\documentclass[onecolumn,10pt]{article}
\usepackage[top=1in, bottom=1in, left=1in, right=1in]{geometry}
\usepackage[numbers,sort&compress]{natbib}
\usepackage{amsmath,amsfonts,amscd,amssymb}
\usepackage{graphicx}
\usepackage{epstopdf}
\usepackage{overpic}
\usepackage{cancel}
\usepackage{rotating}
\usepackage{url}
\usepackage{caption}
\usepackage{color}
\usepackage{rotating}
\usepackage{multirow}
\usepackage{wrapfig}
\usepackage{mathtools}
\usepackage{subeqnarray}
\usepackage{setspace}
\usepackage{palatino} 
\usepackage[bottom,flushmargin,hang,multiple]{footmisc}
\usepackage{lipsum}
\usepackage[font=small,labelfont=bf]{caption}
\usepackage{float}
\newcommand\blfootnote[1]{%
  \begingroup
  \renewcommand\thefootnote{}\footnote{#1}%
  \addtocounter{footnote}{-1}%
  \endgroup
}

\usepackage{lineno}

\definecolor{header1}{cmyk}{0,0,0,1}

\title{\vspace{-.55in}{\fontsize{16}{16}\selectfont \textbf{Go with the FLOW:
Visualizing spatiotemporal dynamics in optical widefield calcium imaging
}}\vspace{-.15in}}

\author{Nathaniel J. Linden$^{1,2}$, Dennis R. Tabuena$^{2,3}$, Nicholas A. Steinmetz$^{4}$,\\ William J. Moody$^{2}$, Steven L. Brunton$^{5}$, Bingni W. Brunton$^{2,*}$\\
\footnotesize{$^1$Department of Bioengineering, University of Washington, Seattle}\\
\footnotesize{$^2$Department of Biology, University of Washington, Seattle}\\
\footnotesize{$^3$Graduate Program in Neuroscience, University of Washington, Seattle} \\
\footnotesize{$^4$Department of Biological Structure, University of Washington, Seattle}\\
\footnotesize{$^5$Department of Mechanical Engineering, University of Washington, Seattle}\\
}
\date{}
\begin{document}
\maketitle

\blfootnote{$^*$ Corresponding author: \url{bbrunton@uw.edu}.}
\begin{abstract}

Widefield calcium imaging has recently emerged as a powerful experimental technique to record coordinated large-scale brain activity.
These measurements present a unique opportunity to characterize spatiotemporally coherent structures that underlie neural activity across many regions of the brain.
In this work, we leverage analytic techniques from fluid dynamics to develop a visualization framework that highlights features of flow across the cortex, mapping wave fronts that may be correlated with behavioral events.
First, we transform the time series of widefield calcium images into time-varying vector fields using optic flow.
Next, we extract concise diagrams summarizing the dynamics, which we refer to as \emph{FLOW (flow lines in optical widefield imaging) portraits}.
These FLOW portraits provide an intuitive map of dynamic calcium activity, including regions of initiation and termination, as well as the direction and extent of activity spread.
To extract these structures, we use the finite-time Lyapunov exponent (FTLE) technique developed to analyze time-varying manifolds in unsteady fluids.
Importantly, our approach captures coherent structures that are poorly represented by traditional modal decomposition techniques.
We demonstrate the application of FLOW portraits on three simple synthetic datasets and two widefield calcium imaging datasets, including cortical waves in the developing mouse and spontaneous cortical activity in an adult mouse.


\end{abstract}

\section{Introduction}

Coordinated organization of neural activity among brain regions is believed to serve many crucial roles, including performing specific computations in the cortex~\cite{wekselblatt2016large,muller2018cortical,musall2019single} and supporting brain development~\cite{corlew2004spontaneous, conhaim2010bimodal,luhmann2016spontaneous}; further, its disruption may lead to neurological disease~\cite{rossi2017focal, cramer2019vivo, mcgirr2017cortical}.
One prominent characteristic of neural activity at the scale of brain regions is the rapid and coherent propagation of activity across cortex, which has been widely observed in a variety of contexts, including spontaneous activity, task engagement, sleep, and development~\cite{siapas1998coordinated,wekselblatt2019distinct,liu2019assessing}.
Qualitatively similar patterns of neural activity propagation have been also observed in the retina, often referred to as retinal waves, during development~\cite{Feller1996requirement, feller1997dynamic, wong1999retinal, tiriac2018light}.
Although such spatiotemporal dynamic features are often visually salient, it remains challenging to quantify and succinctly summarize their behavior directly from neural recordings.

Widefield optical imaging of calcium activity provides a unique opportunity to study coordinated spatiotemporal neural activity among brain areas, because this experimental approach achieves large fields-of-view with high temporal and spatial resolution~\cite{ren2021characterizing, urai2021large}.
In general, widefield imaging experiments involve fluorescence imaging of the entire brain surface of  animals that express optical indicator proteins in known populations of neurons~\cite{sato2012traveling,dana2014thy1,stirman2016wide,silasi2016intact,steinmetz2017aberrant,couto2021chronic}.
Many experiments choose to use genetically encoded calcium indicators from the GCaMP family to image neural calcium dynamics, which is a proxy for electrical neuronal activity~\cite{nakai2001high, tian2009imaging, chen2013ultrasensitive,vanni2014mesoscale}; more generally, the visualization methods we discuss here can be applied to any widefield optical imaging experiment, such as imaging with voltage-sensitive dyes~\cite{mcvea2012voltage,song2018cortical}.
Cortical activity has been measured using widefield calcium imaging in a variety of experiments, notably to study perceptual decision making~\cite{scott2018imaging, musall2019single, wekselblatt2016large, allen2017global, pinto2019task,jacobs2018cortical,zatka2020perceptual},
to extract cortical functional connectivity~\cite{wright2017functional, vanni2017mesoscale,cramer2019vivo},
to characterize cortical activity that organizes brain development~\cite{tabuena2019manuscript},
and to study the effects of disease in the cortex~\cite{cramer2019vivo, rossi2017focal,mcgirr2017cortical}.
In all of these data, it is typical to observe multiple regions activating transiently or in regular succession, with distinct initiation sites and wave-like flows across the fields of view.
These features can often be described as flow of activity with coherent traveling fronts; interestingly, all of these patterns are well studied as nonlinear features of spatiotemporal dynamical systems~\cite{guckenheimer_holmes}.

The most widely applied approaches to analyze time-varying recordings of high-dimensional neural activity are dimensionality reduction techniques, which extract \emph{modes} that correspond to dominant, low-dimensional features of the high-dimensional data~\cite{pang2016dimensionality,cunningham2014dimensionality,dyer2017cryptography}.
These low-dimensional features are useful as representations of the neural activity that facilitate further analysis and modeling.
Furthermore, the observation that the dynamics of neuronal populations can be reduced to a relatively small number of features may be a clue about the mechanisms that underlie coordinated neural activity~\cite{ganguli2012compressed,churchland2012neural, gallego2017neural,gallego2020long}.
Common modal decomposition algorithms used in neuroscience~\cite{cunningham2015linear,pang2016dimensionality} include singular value decomposition (SVD), which is closely related to principle component analysis (PCA), independent component analysis (ICA), and non-negative matrix factorization (NNMF).
These techniques all solve for combinations of relatively few modes in space and time that reconstruct an estimate of the original high-dimensional data; their solutions differ by making different assumptions about the statistical structure of the modes.

\begin{figure}[p!]
    \centering
    \includegraphics[width=0.85\textwidth]{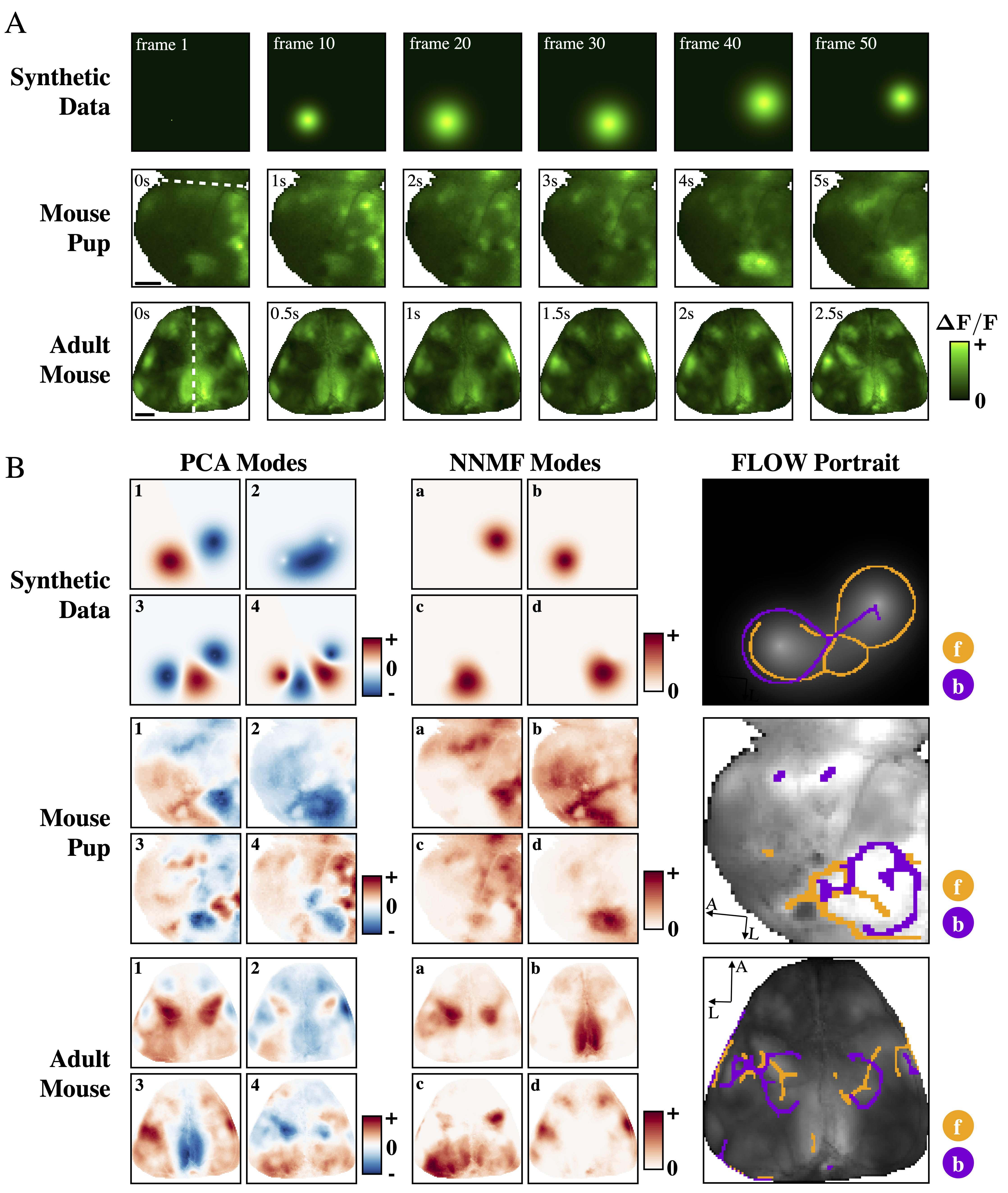}
    \vspace{-.15in}
    \caption{FLOW portraits capture coherent propagation of structures that are poorly represented by common modal decompositions that aim to achieve space-time factorization.
    (\textbf{A}) Three examples of spatiotemporal data for which we compare principal component analysis (PCA), non-negative matrix factorization (NNMF), and our FLOW portraits.
    One synthetic example is a two-dimensional Gaussian that grows, translates to the right, then shrinks.
    Two further \emph{in vivo} examples are widefield calcium imaging data from a developing pup and an adult mouse.
    The dashed white lines at 0 sec indicate the midline of the brain.
    The mouse pup data includes a pan-cortical wave from a postnatal day 7 (P7) animal; scale bar is 1 mm.
    The adult mouse data shows spontaneous widefield calcium activity recorded in the dark; scale bar is 2 mm.
    (\textbf{B})
    FLOW portraits show a succinct summary of the spatiotemporal flow in each example dataset, while spatial PCA and NNMF modes do not.
    The PCA modes are the first 4 spatial components; the NNMF modes are from a 4-mode solution to the factorization and are not ordered.
    Both sets of modes decompose the growth and translation of activity into static spatial images, from which the flow of the activity cannot be easily appreciated.
    In contrast, our FLOW portraits highlight regions of activity initiation and termination, as well as the direction and extent of activity spread. Orange structures ('f'; forward time FTLE) capture regions were activity propagates from, and purple structures ('b'; backward time FTLE) capture regions where activity propagates towards.
    Supplemental videos illustrating all datasets are available as Supplemental Videos 1--5.
    FLOW portraits were computed with integration lengths of 10 frames, 40 frames, and 15 frames and the threshold percentile was set to the 85th, 93rd and 93rd percentiles for the synthetic, mouse pup, and adult mouse datasets, respectively.
    }
    \label{fig:modal_decomp_vs_flow}
\end{figure}

There are many exciting recent innovations in modal decomposition for analyzing large-scale neural data, some of which are extensions and derivatives of SVD, ICA, and NNMF.
Interestingly, while some of these methods have explicit representations of the temporal dynamics (for instance, jPCA~\cite{churchland2012neural}, dynamic mode decomposition (DMD)~\cite{brunton2016extracting,Tu2014jcd,kutz2016dynamic}, and NNMF with temporal constraints~\cite{macdowell2020low,saxena2020localized,mackevicius2019unsupervised,zhou2018efficient}), they largely set out to achieve space/time separation.
Applying PCA and NNMF to segments of synthetic and experimental data (Figure~\ref{fig:modal_decomp_vs_flow}A) yields a set of spatial modes (Figure~\ref{fig:modal_decomp_vs_flow}B; temporal modes not shown) that provide a representation of the activity.
However, these representations are static modes and may not adequately summarize spatiotemporal data.
As an illustrative example, the synthetic data in Figure~\ref{fig:modal_decomp_vs_flow} is a spatial Gaussian that grows, translates, then shrinks with time, and such spatiotemporal coherent features are poorly captured by PCA and NNMF decompositions of the data.

A complementary set of methods have been developed to describe spatiotemporal patterns in widefield neural activity by explicitly extracting propagating waves.
Traveling waves are often characterized by their propagation speed and their direction (see~\cite{zanos2015sensorimotor} and~\cite{blankenship2011role} for examples), and these measures are then aggregated for all of the waves observed in a recording to quantify the trends in wave dynamics.
While this information has proven useful in studying the roles of waves, the approach is limited because waves need to be identified individually.
Several related methods have used the computation of optical flow to convert widefield activity to time-varying vector fields~\cite{afrashteh2017optical, townsend2018detection}.
This velocity field can then be analyzed using tools from vector calculus to identify fixed points, including classifying each fixed point as a source or a sink of activity.
Nevertheless, these methods are constrained to identify only fixed points and cannot identify multiple local planar waves.


Our visualization approach is inspired by the similarity of spatial flows observed in widefield optical imaging to flows of physical fluids.
Humans have a deep intuition about fluid flows from our everyday experiences (e.g., the patterns of milk mixing in coffee, a river flowing).
Representing brain data as a flow allows us to leverage this intuition and decades of methods from flow analysis and visualization.
Propagation of neural activity has many commonalities and differences with physical fluid flows.
In both, there exist coherent structures whose boundaries may be invariant even as the activity changes with time.
In fluid physics, these invariant manifolds are known as Lagrangian coherent structures (LCS)~\cite{Haller2002pof,LCSch3, haller2015lagrangian}, which  act as transport barriers in the flow, either repelling or attracting material.
LCS are often visualized by computing ridges in the finite-time Lyapunov exponent (FTLE) field~\cite{shadden2005definition,Mathur2007prl,Green2007jfm,Brunton2010chaos}, although there are other computational approaches based on variational theory~\cite{Farazmand2012chaos}.
Some noteworthy biological applications include the use of LCS to study the physics of jellyfish feeding~\cite{peng2009transport} and understanding cardiovascular hemodynamics~\cite{duvernois2013lagrangian, shadden2015lagrangian}.
Unlike physical flows, neural activity is not governed by fundamental conservation laws;  nevertheless, these dynamics are well described by time-varying vector fields~\cite{mohajerani2013spontaneous,afrashteh2017optical,townsend2018detection,ashby2019peripheral}.

In this work, we develop a visualization framework to capture the spatiotemporal dynamics of neural activity by extracting field lines in optical widefield imaging, which we call \emph{FLOW (flow lines in optical widefield imaging) portraits}.
FLOW portraits are generated by considering frame-by-frame dynamics as time-varying optical flow vector fields, from which we compute and integrate the ridges in its FTLE.
To validate and develop intuition for our approach, we show that FLOW portraits give accurate and interpretable visual summaries of simple synthetic datasets.
Next, we apply our methods to analyze bouts of activity from two widefield calcium imaging datasets in mice, both of which exhibit spontaneous, widespread activity across cortex.
The first data are recordings of spontaneous cortical activity of GCaMP6s-expressing mouse pups during their first 8 postnatal days~\cite{tabuena2019manuscript}.
The second example is a recording of spontaneous cortical activity in a GCaMP6s-expressing adult mouse~\cite{zatka2020perceptual}.
In both examples, we demonstrate that FLOW portraits extract meaningful and interpretable outlines of the dominant patterns in the cortical activity that contribute to our understanding of the animals'
developmental and behavioral states.

\section{FLOW Portraits}
\begin{figure}[t]
    \centering
    \includegraphics[width=\textwidth]{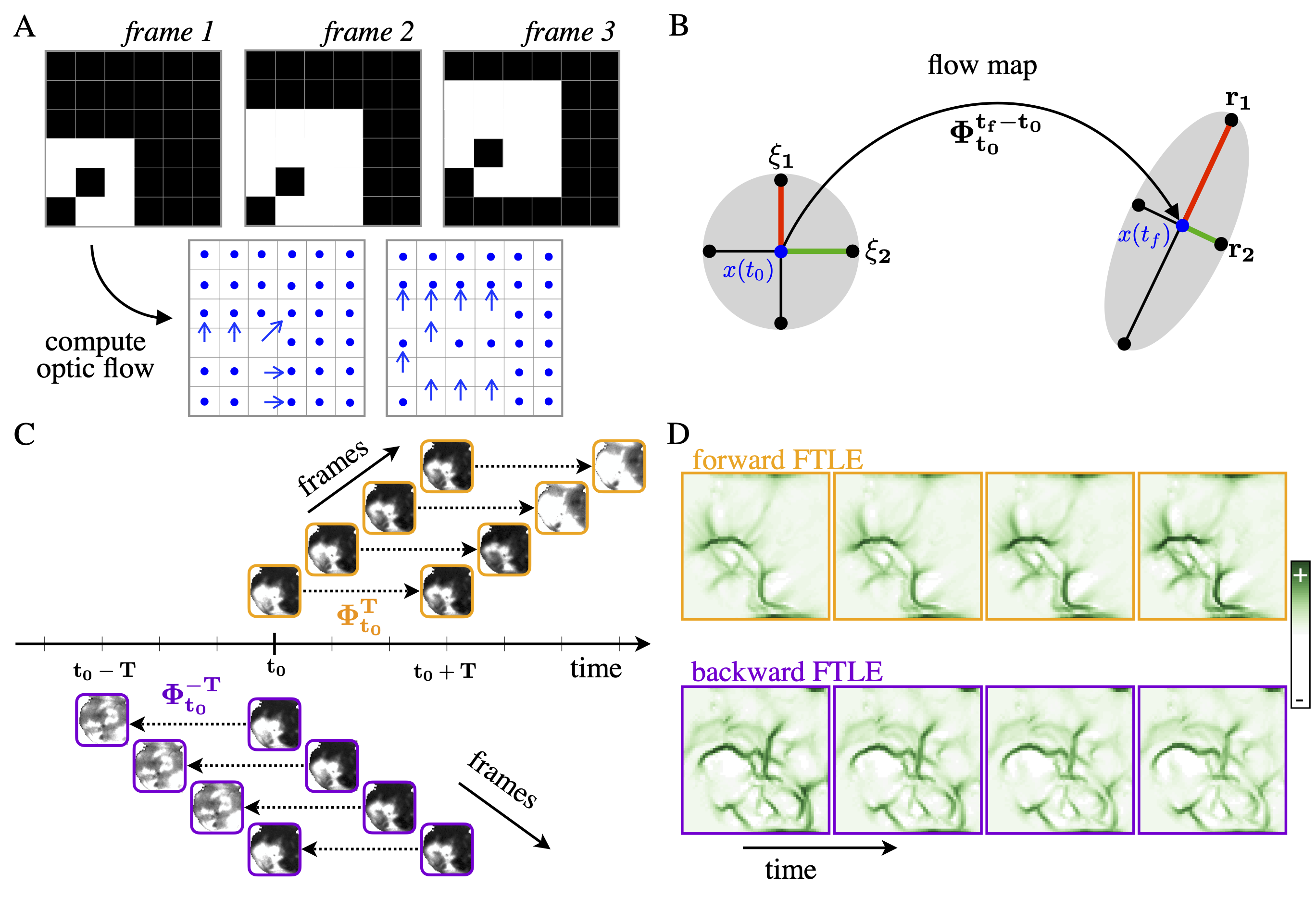}
    \caption{Finite Time Lyapunov Exponent (FTLE) fields are computed from spatiotemporal data.
    (\textbf{A}) An illustration of how optic flow is computed from successive frames of images by correlating the relative movement of pixel intensities. This procedure converts widefield imaging data into a vector field of velocities.
    (\textbf{B}) The flow map at every pixel location is a virtual particle at $x$ integrated through the vector field for a duration of $T$, from $t_0$ to $t_0+T$; in reverse time, particles are integrated from $t_0$ to $t_0-T$. This integration stretches neighboring particles in some directions ($r_1$) and compresses them in others ($r_2$).
    (\textbf{C}) The flow map computation is repeated starting at each frame of the movie, at base time $t_0+k\Delta t$, where $\Delta t$ is the separation between frames; forward maps are orange and backward maps are purple.
    (\textbf{D}) The FTLE fields are computed from the Jacobians of these flow maps; example FTLE fields are illustrated for successive frames of widefield imaging data. }
    \label{fig:ftle_computation_info}
\end{figure}

This work introduces FLOW (flow lines in optical widefield imaging) portraits, which are visualizations that provide a concise and intuitive summary of the spatiotemporal dynamics, highlighting coherent structures in widefield recordings.
Importantly, FLOW portraits differ from modal decomposition techniques in that they do not provide a basis in which to approximate the data and cannot quantitatively explain variance in the recordings.
Instead, FLOW portraits explicitly convert the image stack into time-varying vector fields to extract patterns of activity propagation in the data (Figure~\ref{fig:modal_decomp_vs_flow}).
As our approach leverages and adapts analytic techniques from fluid dynamics~\cite{haller2015lagrangian} that are unfamiliar to most neuroscientists, this section describes how to compute finite time Lyapunov exponent (FTLE) from time-varying vector fields.
We also build intuition for how the ridges of the FTLE field can be interpreted in the context of widefield calcium imaging, using several simple synthetic examples.

The steps of our approach to compute FLOW portraits are illustrated in Figure~\ref{fig:ftle_computation_info},  Figure~\ref{fig:FLOWexample}, and Supplemental Video 1.
The input data is a video (i.e. image stack) of the relative change of fluorescence of the imaged optical protein indicator, ${\Delta F}/{F}$, as it changes in time over many frames.
The raw fluorescence may drift over the course of an experiment, so ${\Delta F}/{F}$ is considered to be a robust proxy for the magnitude of neural activation, normalizing the change in fluorescence over a moving-window baseline~\cite{jia2011vivo}.
FLOW portraits are well suited to summarize data where optical activity is seen to diffuse or flow across the field of view, with varied patterns throughout the recording.
To characterize the propagation of recorded neural activity across brain areas through space, we first compute the flow vector field using optic flow.
Next, the FTLE is computed from the time-varying vector field using the standard integration method as outlined by Onu~\emph{et al.}~\cite{onu2015lcs}.
Last, the FTLE field is post-processed to visualize ridge-like features that highlight the coherent features of a spatial flow~\cite{shadden2005definition, haller2015lagrangian}.
It is important to note that we refer to the processed FTLE ridges as \emph{FLOW portraits} to avoid misinterpretation with traditional LCS analysis in fluid dynamics~\cite{haller2015lagrangian}.
Details of data collection, preprocessing, and computation are described in the Methods (Section~\ref{Sec:Methods}).

\subsection{Optical flow of widefield imaging data}

We describe the frame-by-frame spread of neural activity as time-varying vector fields, computed by optical flow.
Specifically, as regions of high pixel intensity in $\Delta F/F$ move and diffuse across the field of view, these coherent motions can be converted into a vector field of velocities, $dx/dt$ and $dy/dt$, at every pixel in the recording (Figure~\ref{fig:ftle_computation_info}A).
We denote this vector field as $\mathbf{v}(\mathbf{x},t)$, defined at every point in space $\mathbf{x}$ at time $t$.
Motion velocities are commonly estimated from video data in computer vision using optical flow algorithms~\cite{paragios2006handbook}, and biological visual systems of vertebrates and invertebrates also perceive moving scenes with computations akin to optical flow~\cite{duffy1991sensitivity,krapp1996estimation}.
In addition, some prior work has explored optical flow computations in widefield calcium imaging data~\cite{afrashteh2017optical, townsend2018detection}.
Here we use the Horn-Schunck (HS)~\cite{horn1981determining} method because of its simplicity and its observed strong performance on our sample data.

Figure~\ref{fig:FLOWexample}A and B show an example of snapshots of ${\Delta F}/{F}$ data and the extracted optical flow vector fields.
The magnitude and direction of the vector at each pixel is computed by solving for the optimal vector field that describes the change from each frame to the subsequent frame (see schematic in Figure~\ref{fig:ftle_computation_info}A).
In order to  minimize the effects of noise and numerical differentiation on the optical flow field, we apply temporal scaling and smoothing to the computed vector fields.
Briefly, the magnitude of each optical flow vector is scaled proportionally to the relative change in the raw pixel intensity for the corresponding pixel over a prescribed time delay.
This scaling attenuates the magnitudes of vectors that do not represent corresponding changes in the widefield imaging data.
To mitigate the effects of pixel noise, we also apply temporal Gaussian smoothing to the scaled vector fields.
The scaled and smoothed Horn-Schunck optical flow vector fields are used throughout the rest of the FLOW portrait algorithm where velocity data is required.
This process of computing the optical flow vector field from widefield imaging data is analogous to the process of extracting the motion vector field from particle image velocimetery (PIV) data~\cite{willert1991digital, westerweel1997fundamentals} in experimental fluid dynamics.
Both approaches approximate the velocity field from experimental data of \emph{material} transported through the studied flow.

\subsection{The Finite Time Lyapunov Exponent (FTLE)}

Once a flow velocity field, $\mathbf{v}(\mathbf{x},t)$, is computed, there are numerous computational approaches that can be performed to study and characterize the flow.
These methods include instantaneous metrics from vector calculus, such as the divergence and the curl of the vector field, modal decomposition techniques~\cite{Taira2017aiaa,Taira2020aiaa}, such as POD and DMD, and Lagrangian metrics such as the FTLE~\cite{Haller2002pof,shadden2005definition,haller2015lagrangian}.
Although instantaneous metrics have the potential to extract relevant features from widefield imaging optical flow fields (Supplemental Figure 1; \cite{afrashteh2017optical}), the unsteady nature of this data suggests that Lagrangian metrics may provide a more useful summary of the activity.
Here we compute the FTLE fields~\cite{shadden2005definition} to extract time invariant features of flow-like widefield activity.

The FTLE field is a scalar field $\boldsymbol{\sigma}(\mathbf{x},t_0,T)$ defined at every point in space $\mathbf{x}$ and time $t_0$, with respect to some relevant time-scale of integration, $T$.
The FTLE field is a measure of how much neighboring initial conditions separate when integrated through the velocity field $\mathbf{v}$ for a duration $T$.
Thus, regions of high stretching for positive $T$ (forward time) or negative $T$ (backward time) provide time-varying analogs of stable and unstable manifolds, respectively~\cite{guckenheimer_holmes,shadden2005definition,haller2015lagrangian}.
The FTLE field is typically approximated numerically from flow field snapshots at discrete instants in time~\cite{shadden2005definition,Brunton2010chaos}.
First, the flow map $\boldsymbol{\Phi}_{t_0}^{T}$ is approximated on a discretized set of spatial points, typically the same discretized domain where the velocity field is defined.
The flow map $\boldsymbol{\Phi}_{t_0}^{T}$ describes the position of an initial condition $\mathbf{x}(t_0)$ after it is integrated along the vector field $\mathbf{v}$ for a duration $T$ (Figure~\ref{fig:ftle_computation_info}B) and is defined as
\begin{linenomath*}
    \postdisplaypenalty=0
    \begin{align}
        \mathbf{x}(t_0+T) = \boldsymbol{\Phi}_{t_0}^T(\mathbf{x}(t_0)) = \mathbf{x}(t_0) + \int_{t_0}^{t_0+T}\mathbf{v}(\mathbf{x}(\tau),\tau) \,d\tau.
    \end{align}
\end{linenomath*}
Next, the flow map Jacobian $\mathbf{D}\boldsymbol{\Phi}_{t_0}^T$ is approximated via finite-difference derivatives with neighboring points in the flow.
In two-dimensions, the flow map Jacobian at a point $\mathbf{x}$ is:
\begin{linenomath*}
\postdisplaypenalty=0
    \begin{align}
    \renewcommand\arraystretch{2}
    \mathbf{D}\boldsymbol{\Phi}_{t_0}^T(\mathbf{x}) \approx \begin{bmatrix}
    \frac{\boldsymbol{\Phi}_{x,t_0}^T(\mathbf{x}+\Delta x)-\boldsymbol{\Phi}_{x,t_0}^T(\mathbf{x}-\Delta x)}{2\Delta x}
    &
    \frac{\boldsymbol{\Phi}_{x,t_0}^T(\mathbf{x}+\Delta y)-\boldsymbol{\Phi}_{x,t_0}^T(\mathbf{x}-\Delta y)}{2\Delta y}
    \\
    \frac{\boldsymbol{\Phi}_{y,t_0}^T(\mathbf{x}+\Delta x)-\boldsymbol{\Phi}_{y,t_0}^T(\mathbf{x}-\Delta x)}{2\Delta x}
    &
    \frac{\boldsymbol{\Phi}_{y,t_0}^T(\mathbf{x}+\Delta y)-\boldsymbol{\Phi}_{y,t_0}^T(\mathbf{x}-\Delta y)}{2\Delta y}
    \end{bmatrix},
    \end{align}
\end{linenomath*}
where $\boldsymbol{\Phi}_{x,t_0}^T$ denotes the $x$ component of $\boldsymbol{\Phi}_{t_0}^T$, and $\boldsymbol{\Phi}_{y,t_0}^T$ denotes the $y$ component.
The finite-time Lyapunov exponent $\mathbf{\sigma}$ is finally computed from the largest eigenvalue $\lambda_\text{max}$ of the Cauchy-Green deformation tensor $\boldsymbol{\Delta}=\left(\mathbf{D}\boldsymbol{\Phi}_{t_0}^T\right)^\intercal \mathbf{D}\boldsymbol{\Phi}_{t_0}^T$,
which is the maximum singular value of the flow map Jacobian:
\begin{linenomath*}
\postdisplaypenalty=0
    \begin{align}
    \boldsymbol{\sigma}(\mathbf{x}_0,t_0,T) = \frac{1}{|T|}\ln\left(\sqrt{\lambda_\text{max}\left[\boldsymbol{\Delta}(\mathbf{x}_0,t_0,T)\right]}\right).
    \end{align}
\end{linenomath*}
The FTLE value at a point $\mathbf{x}_0$ determines the maximum stretching that may occur between $\mathbf{x}_0$ and a perturbed location $\mathbf{x}_0+\boldsymbol{\epsilon}$ after time $T$:
\begin{linenomath*}
\postdisplaypenalty=0
\begin{align}
    \boldsymbol{\Phi}_{t_0}^T(\mathbf{x}_0+\boldsymbol{\epsilon}) \approx \boldsymbol{\Phi}_{t_0}^T(\mathbf{x}_0) + \mathbf{D}\boldsymbol{\Phi}_{t_0}^T(\mathbf{x}_0)\cdot\boldsymbol{\epsilon},
\end{align}
\end{linenomath*}
where the amplification of the perturbation $\boldsymbol{\epsilon}$ is bounded by
\begin{linenomath*}
\postdisplaypenalty=0
\begin{align}
   \|\mathbf{D}\boldsymbol{\Phi}_{t_0}^T(\mathbf{x}_0)\cdot\boldsymbol{\epsilon}\|_2 \leq \exp(\boldsymbol{\sigma}|T|)\|\boldsymbol{\epsilon}\|_2.
\end{align}
\end{linenomath*}
The $\boldsymbol{\sigma}$ term is understood to depend on $\mathbf{x}_0,t_0,$ and $T$.

The FTLE field is quite robust to noisy measurements of the vector field $\mathbf{v}(\mathbf{x},t)$ ~\cite{Haller2002pof}, since the computation involves integration in time, which tends to average out noise.
This robustness was a major factor in its wide adoption in fluid mechanics, where experimentally acquired velocity fields often contain noise and outliers.
The same robustness is appealing for optical widefield imaging.

\begin{figure}[tb!]
    \centering
    \includegraphics[width=.8\textwidth]{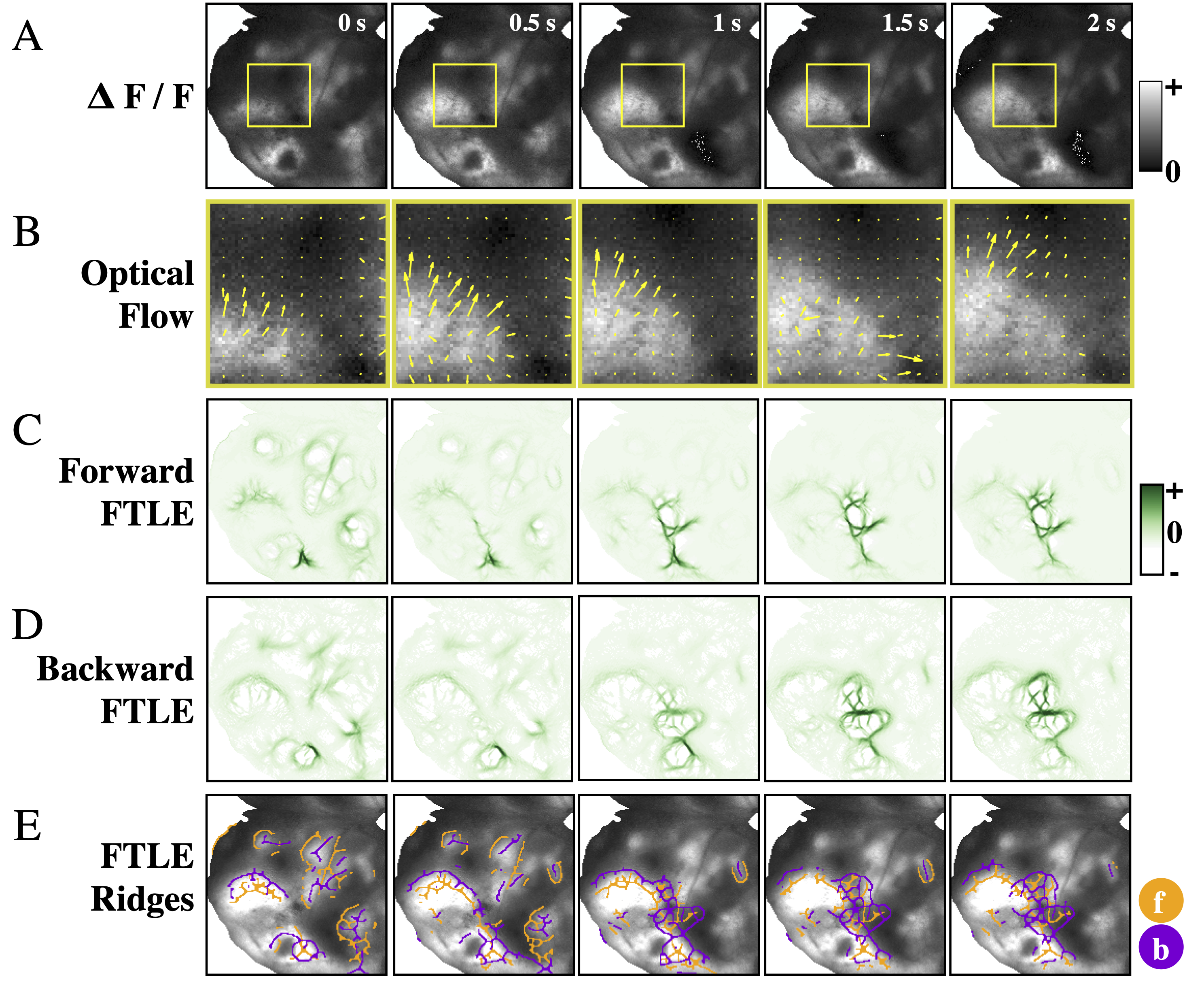}
    \caption{Steps to compute a FLOW portrait.
    Starting with widefield data preprocessed as ${\Delta F}/{F}$ (\textbf{A}), optical flow is used to convert the frame-by-frame changes in pixel intensity to a vector field, shown zoomed in for the smaller area outlined with the yellow box and at $1/6$ spatial resolution for clarity (\textbf{B}).
    Next, the FTLE fields are computed in forwards (\textbf{C}) and backwards (\textbf{D}) time using an integration length of 2 seconds (40 frames); here we show only the non-negative Lyapunov exponents.
    Ridges of these fields highlight coherent structures of the flow (\textbf{E}), and these ridges are used to compute the final FLOW portrait (see Figure~\ref{fig:flow_ridge_extract}).
    The threshold percentile was set to 93 percent.
    The forward time FTLE ridges (orange) highlight regions that repel flow, while the backward time ridges (purple) show regions that attract activity.
    Note that ridges in neighboring frames are similar but do vary in time.}
    \label{fig:FLOWexample}
\end{figure}

Figures \ref{fig:ftle_computation_info} and \ref{fig:FLOWexample} illustrate the intuition behind this FTLE computation, and additional implementation details are provided in the Methods.
The key insight in the FTLE computation is that virtual particles at every pixel location flow according to the vector field from $t_0$ to $t_0+T$, and these integrated optical flow fields form a \emph{flow map} $\boldsymbol{\Phi}_{t_0}^{T}$ (Figure~\ref{fig:ftle_computation_info}B).
This flow stretches neighboring virtual particles, so that equidistant particles have stretched in some directions and compressed in others (see also Supplemental Movie 1).
Relative deformations are described by the Cauchy-Green strain tensor at every pixel, and the FTLE corresponds to the log-normalized leading eigenvalue of this tensor.
The same procedure is repeated by reversing the ordering of frames to compute flow maps in backwards time.
The forward and backward FTLE fields computed for  each example time snapshot are shown in Figure~\ref{fig:FLOWexample}C and D.

Drawing again on our analogy to physical fluid flows, ridges in the FTLE field correspond to time-varying analogs of invariant manifolds, and they approximate   LCS~\cite{LCSch3, haller2015lagrangian}.
In forward time, these features repel fluid material, similar to a \emph{stable} manifold in a dynamical system.
The opposite is true for backward time ridges, where material is attracted in forward time, as with the \emph{unstable} manifold.
A similar interpretation can be extended to the FTLE of optical activity flows, where forward time structures repel activity, while backward time structures attract activity.
However, additional care must be taken when interpreting the intensity of FTLE ridges for brain activity, since the induced velocity field is not divergence free, as is typically the case when analyzing incompressible fluid systems.
When the velocity field is incompressible, then the determinant of the flow map Jacobian is equal to one, so the largest eigenvalue is greater than or equal to one.
However, for compressible vector fields (as is in the case for widefield imaging of neural activity), the divergence is nonzero and the product of the eigenvalues of the flow map Jacobian may not equal to one.
In this case, we may locally have two positive or two negative Lyapunov exponents.
Here we consider only the non-negative Lyapunov exponents, which correspond to repelling ridges in forward time and attractive ridges in backward time (Figure~\ref{fig:FLOWexample}C and D).

\subsection{Ridge extraction for FLOW Portrait visualization}

By aggregating the forward and backward FTLE ridges within a window in time, we summarize the coherent structures of propagating activity within that window with a single FLOW portrait.
Ridges of an FTLE field have been shown to approximate LCS, and several mathematical definitions are suggested to extract them from data~\cite{LCSch3, shadden2005definition, garth2007efficient, lipinski2010ridge}.
We found that implementing existing strategies for ridge extraction on FTLE fields of widefield calcium imaging data did not adequately extract ridge-like features.
Therefore, we developed a post-processing approach to visualize ridges from the forward and backward mean FTLE fields.

\begin{figure}[t]
    \centering
    \includegraphics[width=\textwidth]{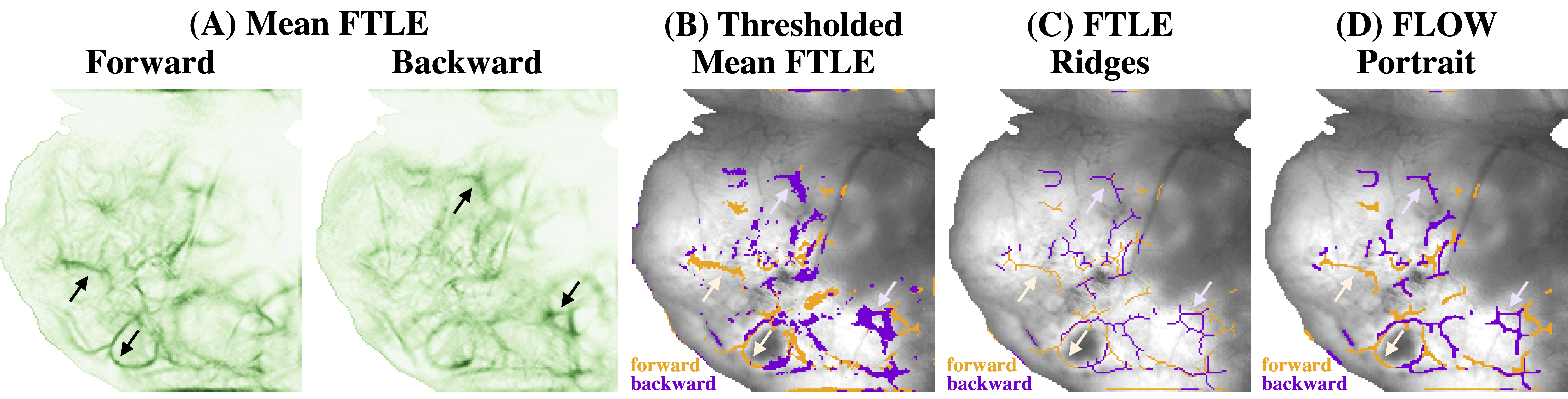}
    \caption{Ridges in the FTLE field are extracted to form the FLOW portrait.
   \textbf{(A)} The forward and backward FTLE fields are separately averaged to aggregate flow structures over time.
   Black arrows indicate examples of FTLE ridges which are extracted in the following analysis.
   \textbf{(B)} Next, the mean FTLE fields are binarized using a threshold which is chosen by the user at a specified percentile (denoted the threshold percentile; here this is chosen to be the 95th percentile).
   The binary forward and backward FTLE fields are shown overlaid on the mean $\Delta F / F$ image.
   Pale orange and purple arrows show the same forward and backwards ridges as in \textbf{(A)}; pale orange corresponds to the forward time ridges and pale purple to those in backward time.
   \textbf{(C)} Ridges in the FTLE are approximated by performing a skeletonization procedure on the binarized FTLE fields.
   \textbf{(D)} Lastly, FLOW portraits are produced by further morphological image processing to smooth the approximate FTLE ridges.
   The final FLOW portrait highlights the example ridges observed in the original mean FTLE fields.
   }
    \label{fig:flow_ridge_extract}
\end{figure}

Ridges lie along local extremes in a field, thus we can approximate their locations by extracting maximal regions and computing the skeleton structure.
To compute the dominant features over the entire recording, we first threshold the mean of all non-negative FTLE values (Figure~\ref{fig:flow_ridge_extract}A) to isolate local maxima in the field (Figure~\ref{fig:flow_ridge_extract}B).
Next, we approximate ridges from the local FTLE maxima by performing a morphological skeletonization operation (Figure~\ref{fig:flow_ridge_extract}C).
Lastly, these ridges are smoothed by applying  morphological image processing  (Figure~\ref{fig:flow_ridge_extract}D).
Thus, the resulting visualization depicts the average approximate FTLE ridges in a recording window to summarize the time invariant patterns of activity.
We refer to this visualization as FLOW portraits because it is designed for compressible vector fields typical of widefield imaging of calcium activity.

There are two parameters the user must choose: the integration time $T$ for the flow map $\boldsymbol{\Phi}_{t_0}^{T}$ and the threshold percentile for FTLE values to include in the visualization.
The choice of these parameters depends on knowledge of the timescales of relevant coherent activity propagation in each dataset.
Larger integration time windows filter out shorter timescale waves; lower percentile thresholds admit more ridges with less intense coherence, which can also admit more spurious ridges if the data are noisy.
Supplementary Figure~4 shows how a range of these parameters yields different FLOW portraits.
As a practical recommendation to users, we recommend repeating the computation for a range of parameter values so that the visually salient features in a dataset are reflected in the FLOW portraits.

\subsection{How to interpret a FLOW portrait}

\begin{figure}[t]
    \centering
    \includegraphics[width=\textwidth]{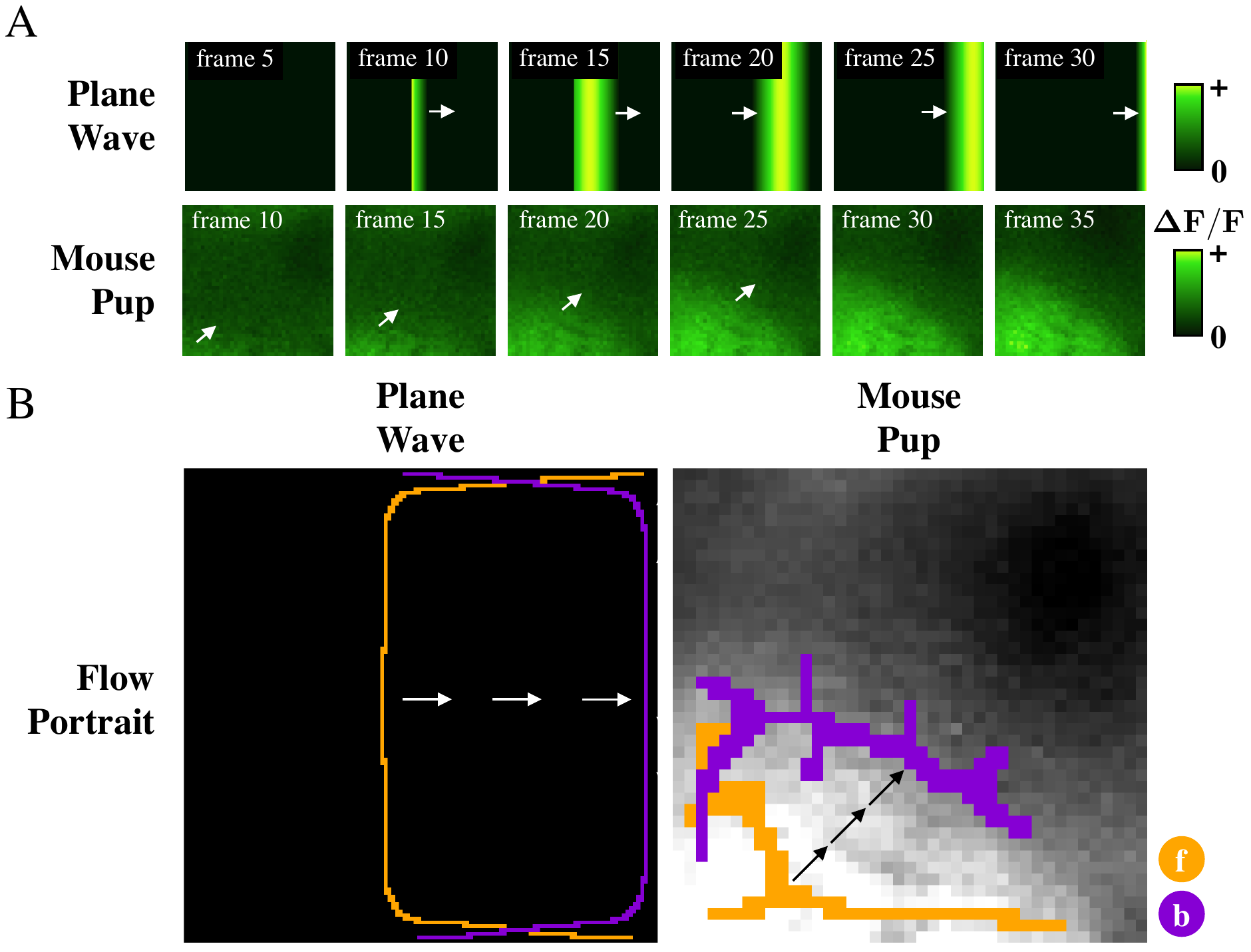}
    \caption{FLOW portraits summarize the activity within a segment of data by highlighting where activity begins and ends.
   \textbf{(A)} Simple traveling waves for which we illustrate FLOW portraits.
   The 'Plane Wave' example shows a synthetic traveling wave which begins in the center of the frame and travels to the right.
   The 'Mouse Pup' example shows a short traveling wave within a segment of widefield calcium imaging experiment of a P7 mouse pup.
   White arrows indicate the direction of activity propagation.
   \textbf{(B)} The FLOW portraits for both waves highlight the regions where the wave begins and were the wave ends.
   Arrows (black and white) show the general direction of wave propagation.
   FLOW portraits were computed with integration lengths of 15 frames and 5 frames and the threshold percentile was set to the 91st and the 90th percentiles for the plane wave and mouse pup datasets, respectively.}

    \label{fig:plane_wave_examples}
\end{figure}

To build intuition and illustrate how spatiotemporal patterns are visualized by FLOW portraits, let's examine them for several simple synthetic datasets, each capturing the types of coherent activity commonly observed in widefield calcium imaging.
The first example is a plane wave that starts in the middle of the field-of-view and travels to the right (Figure~\ref{fig:plane_wave_examples}A).
In the corresponding FLOW portrait, the forward-time FTLE structures delineate where the wave originates in the middle of the field-of-view, while the backward-time FLTE structures outline where the wave terminates (Figure~\ref{fig:plane_wave_examples}B).
This type of traveling plane wave closely resembles the spread of neural activity observed by widefield imaging (for instance, data from mouse pup in Figure~\ref{fig:plane_wave_examples}).
The second synthetic dataset is a circular wave that initiates in the middle, then grows larger towards the edges (Supplemental Figure 2).
Here, the forward-time FTLE structures mark the site of initiation, while the backward-time FTLE structures outline the maximal spatial extent of the circle's spread.
Our third synthetic example combines both traveling and growing/shrinking wave fronts.
As shown in Figure~\ref{fig:modal_decomp_vs_flow}, Supplemental Video~1, and Supplemental Figure~2, the 2D Gaussian dataset includes a Gaussian blob that appears in the field-of-view, grows in diameter, translates to the right, then shrinks.
Forward-time FLTE structures capture where the activity originates, including the back edge of the Gaussian as it starts to translate and the outside perimeter of the blob as it shrinks.
Similarly, backward-time FTLE structures capture where the activity terminates, including the outside perimeter of the blob as it grows and the center of the shrinking blob.

In all of these examples, FLOW portraits represent succinct summaries of spatiotemporal coherent activity, highlighting regions of activity initiation and termination, as well as the direction and spatial extent of how activity spreads.
Specifically, activity originates \emph{from} the forward-time FLOW ridges (orange lines, analogous to stable manifolds) and goes \emph{to} the backward-time FLOW ridges (purple lines, analogous to unstable manifolds).
This visualization caricaturizes features of coherent activity not accessible by established methods, including modal decomposition (Figure~\ref{fig:modal_decomp_vs_flow}), instantaneous metrics like divergence and curl (Supplemental Figure 1), and source/sink classification of fixed points (Supplemental Figures 2 and 3).
The forward and backward time FTLE structures carry more information than sources and sinks because they are not constrained to be fixed points; thus, these structures are able to delineate traveling fronts.
The intersection of two or more FLOW structures, such as where the orange and purple ridges intersect in Figure~\ref{fig:modal_decomp_vs_flow}B, can occur for several reasons.
First, intersections of the forward and backward FTLE ridges are reflected as intersections in the FLOW portraits.
Points where these FTLE ridges intersect correspond to time-dependent saddle points, as the forward and backward FTLE ridges are time-dependent analogs of the stable and unstable manifolds of the vector field.
Second, two different spatiotemporal structures may occur at the same spatial location at different times during the recording, as in the case of the 2D Gaussian synthetic dataset.

\section{FLOW portraits of widefield calcium imaging data}

We demonstrate the application of our approach on several optical widefield datasets, all recordings of spontaneous calcium activation imaged from the cortical surface of transgenic mice.
In each example, we have chosen to focus on windows in time when bouts of activity are observed across large portions of cortex.
We show that FLOW portraits extracted from these windows summarize the extent and direction of calcium flow, highlighting cortical areas whose neural activations can be interpreted in the context of the behavioral and developmental context of the animals.

\begin{figure}[htp]
    \includegraphics[width=\textwidth]{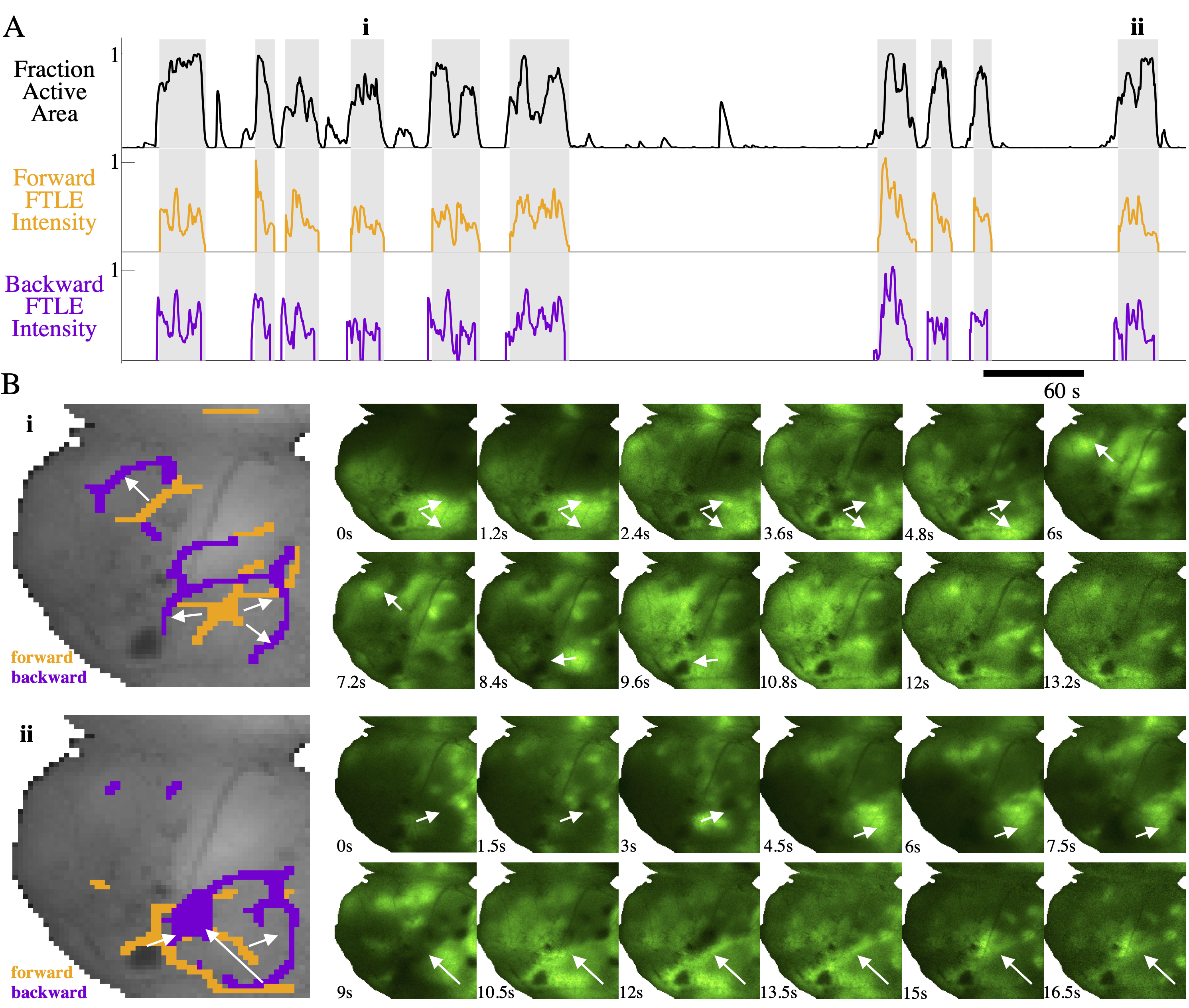}
    \caption{Pan-cortical wave events in a P7 mouse pup are summarized as FLOW portraits.
    (\textbf{A}) Pan cortical waves are defined as events where the fraction of active cortex (black-trace) exceeds 50-percent.
    Briefly, fraction of active cortex is defined as the fraction of pixels whose intensity is greater than one standard deviation above the mean (in time) for that pixel.
    FTLE intensity is defined as the sum of the FTLE values for each frame, normalized by the maximum value in time; this intensity is computed for both the forward and backward FTLE time series.
    (\textbf{B}) FLOW portraits are shown for two example waves, indicated by \textbf{i} and \textbf{ii} in \textbf{A}.
    Orange indicates forward time FTLE ridges where calcium activity originates.
    Purple indicates backward time FTLE ridges where calcium activity propagates towards.
    White arrows highlight the general direction of activity propagation during the cortical wave.
    The FLOW portraits are computed using an integration length of 2 seconds (40 frames) and a threshold percentile of 93 percent.}
    \label{fig:moody_cortical_waves}
\end{figure}

\subsection{Example 1: Pan-cortical waves}

Pan-cortical waves are bouts of activity that propagate across large areas of the cortex~\cite{garaschuk2000large,conhaim2010bimodal, conhaim2011developmental, easton2014genetic, barger2016early} and are suggested to play a critical role in cortical development~\cite{tabuena2019manuscript}.
These events are defined heuristically as activity that propagates to include a large fraction of the imaged cortical surface.
In Figure~\ref{fig:moody_cortical_waves}A, the gray bars highlight individual cortical wave events, defined as when the fraction of active cortex rises to above $1/2$ and falls back to the baseline ($\sim1/10$).
To contribute to our understanding of pan-cortical waves in development, we use FLOW portraits to summarize the activity during each wave event, thus facilitating direct comparisons across individual waves and developmental time points.

We construct FLOW portraits to summarize the flow of activity during each pan-cortical wave.
Spatial integration of the FTLE fields yields the FTLE intensity (Figure~\ref{fig:moody_cortical_waves}A, orange and purple traces), which indicates the relative amount of time-averaged \emph{flow} throughout the recording.
The resulting FLOW portraits for two pan-cortical waves can be seen in Figure~\ref{fig:moody_cortical_waves}B, alongside 12 frames of the ${\Delta F}/{F}$ data from each wave (see also Supplemental Movies 2 and 3).
The portraits of every pan-cortical wave are shown in Supplemental Figure~5.

Each FLOW portrait provides a summary of the prominent activity observed during each wave event, highlighting the regions that repel (forward FTLE, orange) and attract (backward FTLE, purple) activity.
Indeed, both waves shown in Figure~\ref{fig:moody_cortical_waves}B exhibit two stages of propagation, where activity spreads and pauses briefly at sensorimotor cortex (outlined by the purple rings) before spreading towards frontal cortex.
This concise visualization allows us to easily compare such qualitative features of wave propagation without having to parse through the raw data manually.

\subsection{Example 2: Sleep-state cortical activity changes in development}
\begin{figure}[t!]
    \centering
    \includegraphics[width=\textwidth]{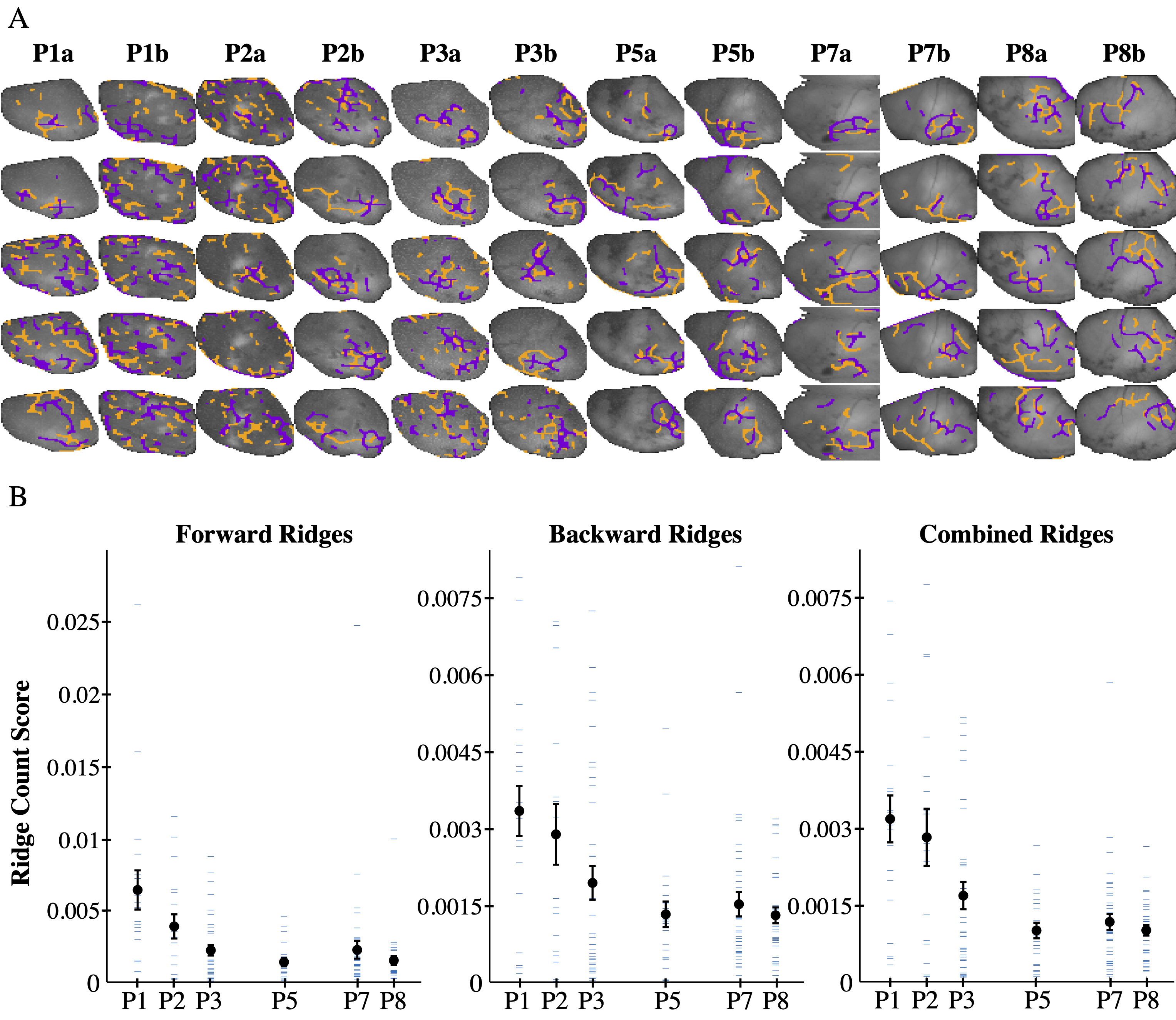}
    \caption{FLOW portraits highlight developmental changes of sleep-state cortical activity.
    (\textbf{A}) FLOW portraits for 5 sleep bouts from 12 P 1-8 mouse pups are shown.
    During the first 3 postnatal days, activity is diffuse, as indicated by many short-length structures in the FLOW portrait.
    As animals grow older (postnatal days 5-8), sleep-state cortical activity becomes more structured, as indicated by a consolidation of features in the FLOW portraits.
    Orange indicates repelling structures, and purple indicates attracting structures.
    All images are of the left-hemisphere, such that the mid-line and anterior directions are oriented towards the bottom and left of the images, respectively.
    FLOW portraits were computed using an integration length of 2 seconds (40 frames) and a threshold percentile of 93 percent for all sleep-bouts shown.
    (\textbf{B}) Quantification of the number of FLOW ridges seen versus developmental day.
    The ridge count score for a FLOW portrait is computed by counting the number of FLOW ridges, either forward, backward, or both combined, and dividing by the total area of FLOW ridges in that portrait.
    When the ridge count score is high there are many smaller ridges in the image, whereas when the score is low there are fewer ridges with a larger ridge area.
    Here, the mean ridge count score (black point) decreases between developmental day one and day five and then remains constant.
    Further, the mean ridge count score over days P1-P3 ($0.0038$, $0.0027$, and $0.0022$ for forward, backward and combined, respectively) is significantly different than the mean ridge count score over day P5-P8 ($0.0019$, $0.0015$, and $0.0010$ for forward, backward and combined  respectively; paired t-test, p-values $8.70\times10^{-4}$, $7.16\times10^{-5}$, and $1.62\times10^{-7}$ for forward, backward and combined, respectively).
    Blue dashes show individual data points; the ridge count score for an individual FLOW portrait.
    Error bars show $\pm 1$ standard error measure.}
    \label{fig:moody_sleep}
\end{figure}

To further investigate the role of spontaneous cortical activity during development, we analyzed optical recordings of spontaneous calcium activity in mouse pups during the first 8 postnatal days of development.
We computed FLOW portraits on bouts of spontaneous cortical activity during sleep in 12 animals of ages P1, P2, P3, P5, P7 and P8 (Figure~\ref{fig:moody_sleep}).
Briefly, the sleep state was determined by binning time points into three categories (sleep, wake, and moving-wake) using the power of nuchal EMG spectrum~\cite{seelke2010developmental, blumberg2014development, tabuena2019manuscript}.
We chose to focus on sleep state cortical activity for its proposed developmental roles and observed changes during development~\cite{tabuena2019manuscript}.
For each animal, we computed FLOW portraits for up to the 10 longest bouts of sleep (fewer portraits were computed for short recordings where there were less than 10 sleep bouts).

Five example FLOW portraits for each animal are in Figure~\ref{fig:moody_sleep}A, with a complete set in Supplemental Figure~6. 
This organization allows us to leverage FLOW portraits to examine developmental changes in cortical activity across long recordings from different animals.
We observe a qualitative change between the portraits from the early postnatal days (P1--3) to the later days (P5--8).

The FLOW portraits from the early days show more diffuse activity, with less consolidated FTLE ridges.
After P5, the FLOW portraits show cortical activity during sleep becoming more consolidated and following more defined flow patterns.
We quantify this transition to more consolidated FTLE ridges after P5 by defining a ridge count score.
Briefly, this metric is computed by counting the total number of disconnected ridges in a FLOW portrait and dividing this sum by the total area of FLOW portrait.
We computed ridge count scores for all FLOW portraits seen in Supplemental Figure 6 and summarized the mean over each developmental day (Figure~\ref{fig:moody_sleep}B).
We found that the ridge count score for forward FTLE structures, backward FTLE structures, and both combined all decreased between P1 and P5.
Further, we found that the mean ridge count score over the early developmental days (P1--P3) was significantly different (p-values of $8.70 \times 10^{-4}$, $7.16 \times 10^{-5}$, and $1.62 \times 10^{-7}$ for forward, backward, and combined, respectively; paired t-test) than that over the later developmental days (P5--P8).

\subsection{Example 3: Cortical activity during spontaneous movement}
Lastly, we analyze the FLOW portraits of spontaneous cortical activity in a head-fixed, behaving adult mouse~\cite{zatka2020perceptual}.
To investigate how FLOW portraits align with an animal's behavior, we analyze IR videos of spontaneous facial and limb movements alongside cortical calcium activity.
A movement score was assigned to each recording time point by using the total pixel-wise difference between the current and next frames (the forward difference) and normalizing this to the maximum observed difference.
During bouts of limb movement or whisking the movement score was greater, approaching the maximum score of 1, than during periods of rest, when the score approached the minimum score of 0.

\begin{figure}[htp]
    \centering
    \includegraphics[width=\textwidth]{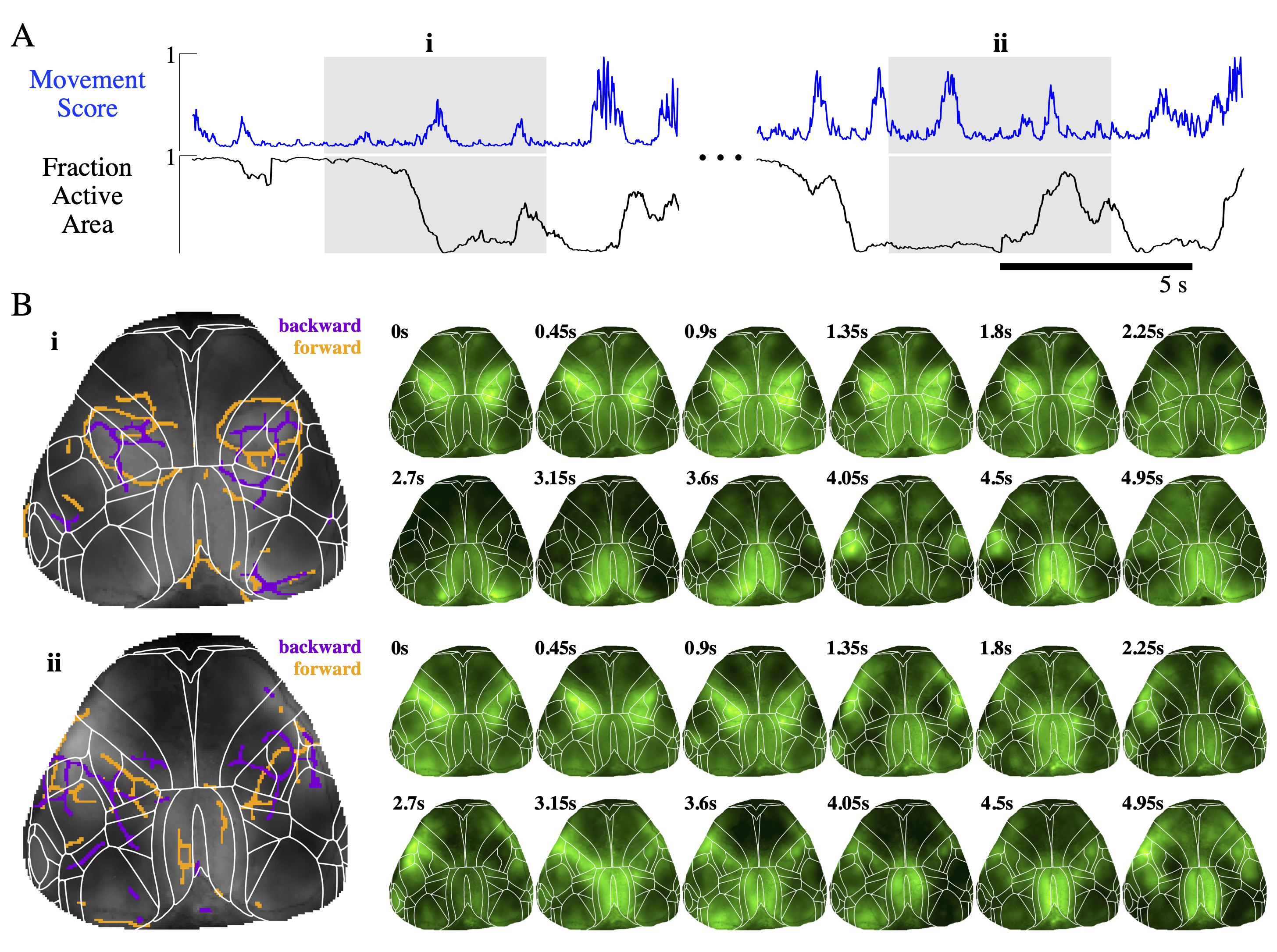}
    \caption{Examples of spontaneous cortical calcium activity associated with movements of an adult mouse summarized as FLOW portraits.
    (\textbf{A}) A movement score extracted from IR video of the mouse moving spontaneously in the dark shows bouts of large movements among more quiescent periods.
    These bouts of movements do not correspond necessarily to when a large fraction of the cortical surface is active (see Methods for threshold criteria).
    (\textbf{B}) FLOW portraits for two bouts involving spontaneous movements labeled \textbf{i} and \textbf{ii} show coherent structures that highlight activity appear in sensorimotor regions and are then attracted to the centers of these regions bilaterally.
    Boundaries aligned to the Allen Mouse Brain Common Coordinate Framework~\cite{wang2020allen} are overlaid in white.
    FLOW portraits were computed with a 15 frame integration length and threshold percentile of 93 percent.}
    \label{fig:steinmetz_movement}
\end{figure}

We chose two bouts of spontaneous movement (gray shading in Figure~\ref{fig:steinmetz_movement}A highlights the two bouts, \textbf{i} and \textbf{ii}) to compute the corresponding FLOW portraits (see also Supplemental Movies 4 and 5).
Large variations in the movement score (Figure~\ref{fig:steinmetz_movement}A, blue trace) can be observed throughout these bouts, indicating that the animal is continuously switching from a resting to a moving state.

We see signatures of these movement behaviors in the calcium activity, when we expect sensorimotor cortical regions to be more active than during periods of rest.
Indeed, the FLOW portrait for each activity bout provides a clear summary of calcium activity surrounding the sensorimotor cortex (Figure~\ref{fig:steinmetz_movement}B).
During both bouts, a ring-like repelling (forward, orange) field line outlines the sensory-motor region, while attracting (backward, purple) field lines fill in the centers of the rings.
We note that these patterns are different from our analysis of Example 1 of pan-cortical waves.
Specifically, these features suggest a dominant pattern of cortical calcium activity as diffusion of activity from the entirety (or outer edges) of sensorimotor regions towards the center.
In other words, our FLOW portraits point to sensory-motor cortex as a terminus of cortical activity during spontaneous movement behaviors.
Interestingly, compared to the overlaid Allen Mouse Brain Common Coordinate Framework (white lines in Figure~\ref{fig:steinmetz_movement}B), the attracting (backward, purple) field lines are close to the boundary between somatosensory and primary motor cortices.
We note that the integration length and the threshold percentile parameters chosen for these examples determine which ridges are highlighted in the FLOW portraits.

\section{Discussion}

This paper introduces FLOW portraits as a novel approach to visualize spatiotemporal flow of coherent features in optical widefield calcium imaging data.
Viewed at this meso-scale of temporal and spatial resolution, neural activity at the cortical surface is typified by multiple brain regions activating transiently and sometimes in spatial succession.
Motivated by an analogy between this flow of neural activity over cortex and physical fluid flows, we leverage techniques well established to study physical fluid flows, in particular the finite-time Lyapunov exponent (FTLE).
Here we convert movies of $\Delta F/F$ over the cortical surface into vector fields, and the FTLE ridges in these vector fields form an intuitive map of dynamic calcium activity.
Importantly, our FLOW portraits do not decompose the data into modes and are not models of the data.
Instead, they capture succinct portraits of diverse, variable, and non-stationary spatiotemporal patterns, such as those often observed in spontaneous or task-driven widefield calcium imaging experiments.

The FLOW portrait analysis makes several assumptions that are usually true of physical fluid systems but often not met by neural data.
Coherent propagation of neural activity on the cortex does not obey mass or energy conservation, so the extraction of FTLE ridges are only approximate ``material'' accumulation lines.
This assumption is particularly invalid for long bouts of data and over long integration windows, so caution must be exercised in choosing these parameters in the analysis (the same is true of FTLE analysis in fluid flows).
Because the integration window effectively low-pass filters the dynamics of the data, activity that is on a faster timescale may be attenuated, and local activity may integrate to appear more coherent.
The optimal choice of FTLE parameters for visualization widefield activity and how these depend on spatiotemporal statistics will be important to understand in future applications.
Further, although widefield imaging offers much larger fields of view at a higher temporal resolution than many other imaging methods, there remains much unobservable neural activity.
Brain areas outside the imaging window and underneath the cortical surface contribute to the imaged activity, yet the flow of neural activity among these regions cannot be captured by our analysis and may bias the extracted flow lines.
This limitation is more severe in considering brains with sulci and gyri, as our analysis fundamentally assumes that neighboring pixels are also neighbors on the cortical sheet.

The quality and interpretability of FLOW portraits requires the imaging data to have been acquired with sufficient temporal and spatial resolution to support the analysis.
To be specific, we require that the sampling in time be fast enough that successive frames of the movie are very similar.
If the frame rate is too slow and neighboring frames differ substantially, then the optical flow computation infers inaccurate vector fields and can no longer disambiguate between gradual flow of activity and sudden jumps in activation.
Despite the relatively slow dynamics of GCaMP6s compared to single neuron activity~\cite{chen2013ultrasensitive}, the temporal dynamics of lasting neural synchrony at this meso-scale is adequately matched to the kinetics of the indicator protein in all the data we highlight here.
The choice of calcium or voltage indicator also introduces filtering in time, so that our analysis relies on the dynamics of the indicator to be faster than the dynamics of the underlying flow across the brain.
Similarly, the spatial resolution of the data need not support disambiguation of single neurons, but it is important that spatial averaging in the field of view does not obscure coherent features of interest.

We suggest our approach expands the toolbox of techniques to analyze and understand widefield imaging data, especially facilitating direct comparison of multiple bouts of spatiotemporal activity that are interpretable in the context of behavior and development.
This visualization framework can be developed to explicitly quantify features of the flow (for example, the ridge count score analysis in Figure~\ref{fig:moody_sleep}).
Such quantification may be of value in further work that connects features of FLOW portraits with states of relevance to behavior, development, or disease.
The transformation of widefield calcium imaging data into a vector field representation suggests multiple avenues for development of analytic tools.
For instance, where multiple coherent waves are present and propagate locally, future work may develop visualizations of the direction of activity propagation, from individual forward FLOW ridges to backward FLOW ridges.
Intriguingly, it may be possible to discover partial differential equations that govern the flow of activity through these vector fields using data-driven techniques~\cite{rudy2017data,Schaeffer2017prsa}.

\section{Methods}\label{Sec:Methods}

\subsection{Widefield calcium imaging and data preprocessing}

\subsubsection{Developing mouse datasets}

These experimental procedures were conducted at University of Washington, and all protocols were reviewed and approved by the University of Washington IACUC.
Neonatal mice expressing GCaAMP6S in cortical neurons were bred by crossing mice heterozygous expressing a Emx1 driven Cre (Emx1-Cre+/-, Jackson Labs ID 005628) with mice homozygously expressing GCaAMP6S under control of a cre promoter (Ai162+/+, Donated by Allen Institute, Jackson Labs ID 031562).
This cross resulted in mice expressing GCaAMP6S primarily in glutamatergic cortical neurons early in development.
On the day of recording, mice were placed on a heating pad and anesthetized using 1--2\% isoflurane carried by 100\% O2, while local anesthetic bupivacaine was delivered subcutaneously at the scalp.
The skin over the cortex was removed over a window spanning between the ears to just above the eyes of the pup, to reveal the skull.
The periosteum was then removed with fine tip forceps and cotton swabs.
At this developmental stage, the skull is uncalcified and largely transparent, so thinning or cutting a window was unnecessary.
A stainless steel U-shaped bracket was then attached to the skull with cyanoacrylate glue.
The bracket was clamped in place to the heating pad and stage to stabilise the head.
To prevent the skull from drying and to preserve clarity, the exposed skull was also sealed with a thin layer of cyanoacrylate.
Silver wire hook leads were implanted in to the nuchal muscle through the same incision to monitor neck electromyography (EMG).

Once glue had dried, isoflurane anesthesia was removed and the pup along with heating pad and stage was positioned for imaging on a Nikon AZ100 with 2X objective and 0.6X reducer.
Nuchal EMG activity was amplified with and AM Systems Model 1700 amplifier (10Hz high pass, 60Hz notch, 10kHz low pass) and was sampled at 10kHz using a Powerlab 4/26 and Labchart v8 (AD Instruments).
GCaMPP6s activity was excited using an Intensilight mercury lamp (Nikon), captured using an CCD camera (ORCA Flash 2.8), and recorded using the HCImage application (Hamamatsu).
Frame capture rates varied from 10--50Hz with maximum exposure times (100--20ms, respectively).
To further increase signal to noise ratio, the camera was set to perform online hardware based pixel binning, reducing a 1920$\times$1440p image to 960$\times$720p.
Individual recordings began when the animal began cycling regularly between sleep and wake, and recordings typically lasted between 40--60 minutes, after which the pup was euthanized.

Ca2+ records were processed using MATLAB (Mathworks) to create ${\Delta F}/{F}$ image stacks for FLOW portrait analysis.
Briefly, imaging runs were further downsampled by pixel binning the 960$\times$720p image down to 480$\times$360p.
To compensate for slow drift, a moving window of 40-sec was used to calculate baseline $F$ for each frame; each pixel in $F$ was set to the minimum value for that pixel across the 40-sec window.
${\Delta F}$ was calculated as the difference between the raw pixel intensity and this calculated moving minimum.
The difference was then normalized to relative change by dividing (${\Delta F}/{F}$).
A small Gaussian spatial blur was used to attenuate ``speckled'' noise.
Region of interests (ROI) masks of the visible cortical surface were generated by excluding any pixel whose mean-to-variance ratio was greater than 400:1.
This value was determined heuristically to optimize exclusion of any pixels that displayed minimal change in fluorescence over time, such as those that lie outside the cortical window.

\subsubsection{Adult mouse dataset}
These experimental procedures were conducted at UCL according to the UK Animals Scientific Procedures Act (1986) and under personal and project licenses granted by the Home Office following appropriate ethics review.
The dataset and associated procedures was described previously ~\cite{zatka2020perceptual}.
In brief, the data were from an adult (30 weeks) male mouse expressed GCaMP6s in excitatory neurons (tetO-GCaMP6s; CaMK2a-tTa genotype~\cite{wekselblatt2019distinct}).
The mouse was implanted with a metal headplate, plastic light isolation chamber, and transparent covering over the dorsal skull.
On the day of recording, the mouse was head-fixed under the microscope on a stable seat with a rubber wheel underneath the forelimbs.
Video cameras captured the frontal aspect of the mouse as well as its eye. Imaging was conducted at 70 Hz with alternating blue and violet illumination, and imaging data was corrected for hemodynamic components. The data were processed by singular value decomposition (SVD) compression.

The images were aligned to the Allen Common Coordinate Framework~\cite{wang2020allen} by manually identifying Bregma and the orientation of the midline in the images.
Bregma was taken to be located at the coordinate 5.7  mm AP in the CCF.
Since the pixel size in the camera was known (21.7  $\mu m$ / pixel), the CCF region boundaries could then be overlaid on the images.
\subsection{Imaging analysis}
\subsubsection{Pan-cortical wave segmentation}
Pan-cortical waves, as defined by~\cite{tabuena2019manuscript}, are cortical activity events where recorded activity spreads over a large area of the imaged cortex.
We defined large cortical area to be when 50 percent of the cortical pixels (pixels which show the cortex) are active.
At any time point, a pixel is active if its intensity is more than one standard deviation above the temporal mean for that pixel.
To extract pan-cortical wave events, we computed the fraction of active cortical pixels throughout the recording, and noted the time points where the active area exceeded the 50 percent threshold.
Each wave event was then defined by the time points when the active area crossed 10 percent active prior to the time of crossing the 50 percent threshold and the time when the active area crossed this 10 percent lower bound following the peak.
Overlapping wave events were merged into a signal pan-cortical wave to avoid redundancy.
Furthermore, events that lasted less than the FTLE integration length ($T$) plus the optical flow scaling delay ($3.5$ sec or $70$ frames for the mouse pup data) were not analyzed because the FTLE and optical flow computations require longer bouts of data.
\subsubsection{Sleep bouts during development}
Sleep state cortical activity was segmented using the nuchal EMG as an indicator of state (sleep or awake).
Time points were clustered into three groups based on the nuchal EMG power spectrum as in~\cite{blumberg2014development, seelke2010developmental, tabuena2019manuscript}, where the lowest power group is known to represent the sleep state.
We defined a sleep bout as a period of continuous classification in the sleep state, and extracted the 10 longest bouts from each recording over the developmental time span.
Any bout that did not meet the FTLE and optical flow length requirement ($3.5$ sec or $70$ frames for the mouse pup data; $0.8$ sec or $30 $frames for the adult mouse data) was not analyzed further.
In cases when there were less than 10 bouts that met the length requirement, we chose to include fewer sleep bouts for that recording.
\subsubsection{Movement event extraction}
We extracted movement events from video of the face and front arms of the adult mouse during the widefield imaging experiment.
We defined a movement score for each time point in the video based on the difference between the current time point and the previous time point.
Each video frame was assigned a movement score given by the sum (over all pixels in the frame) of the difference between the current and previous frame.
For time point $t$, the score is given by $\text{MovementScore}_t = \sum_\text{pixels}{(I_t - I_{t-1})}$, where $I$ is the pixel intensity for each of the pixels in the frame.
The time series of movement scores was normalized to the maximum observed value for ease of interpretation and visualization.
Timestamps of video frames were determined by recording TTL pulses emitted by the camera on each exposure, for both calcium imaging and behavioral videos.
We then compared cortical activity across varying movement regimes.
\subsection{Optical flow computation}
\subsubsection{Horn-Schunck optical flow}
We computed optical flow vector fields using the Horn-Schunck optical flow algorithm~\cite{horn1981determining} implemented in MATLAB~\cite{hsofcode}.
Two parameters must be supplied to the optical flow algorithm: the maximum number of iterations and the $\alpha$ smoothness parameter.
Values for both parameters were selected such that the errors in the Horn-Schucnk minimization problem (see~\cite{horn1981determining} for details) were simultaneously minimized.
We set the maximum number of iterations to 100 and $\alpha$ to $1$ for all computations.
\subsubsection{Optical flow scaling and smoothing}
To minimize the effects noise on optical flow fields, we applied an activity-based scaling to the magnitudes of the optical flow vectors.
First, we created a time series of weights for each pixel by normalizing change in raw pixel intensity between the current time and the intensity of that pixel $1.5$ seconds in the past to the maximum observed change.
We chose a time delay of $1.5$ and $0.5$ seconds, for the developmental and adult mouse datasets respectively, to empirically to match the time scale of large changes observed in the raw data.
Next, we took the sliding windowed average, over a window of $0.25$  seconds, of the weights in order to further reduce the effects of recording noise.
We then scaled the magnitude of the optical flow vectors by applying the weights to the corresponding vector.
Lastly, we temporally smoothed the optical flow fields using a 5-point Gaussian window created with MATLAB's \verb|guasswin()| function.
The \verb|gausswin| function takes an additional parameter, $\alpha$, which is proportional to the inverse of the standard deviation of the Gaussian smoothing kernel.
We set this parameter to $1.25$ for all smoothing operations for it observed ability to reduce noise in the processed vector fields.
\subsection{Finite Time Lyapunov Exponent (FTLE) fields}

We computed the FTLE of all vector fields using the LCS Tool~\cite{onu2015lcs} (\url{https://github.com/jeixav/LCS-Tool}) MATLAB software package.
We computed the FTLE using an integration length of $2.0$ seconds ($40$ frames) for the developing mouse data and an integration length of $\sim0.5$ seconds ($15$ frames) for the adult mouse data.
Additionally, we used an integration lengths of 15 frames, 12 frames, and 10 frames for the plane wave, the circular wave, and the traveling Gaussian examples respectively.
To chose the integration length $T$, we followed the criteria outlined in~\cite{shadden2005definition} of choosing a value such that the FTLE ridges are sufficiently resolved.
Using a sample of each dataset, we computed the FTLE for a range of integration lengths ($0$ to $100$  frames) and visualized the resulting FTLE fields.
We then chose the smallest integration length where the corresponding FTLE field had well resolved, sharp, ridges.
Supplemental Figure 4B illustrates the effects of computing FLOW portraits with a range of integration lengths.
\subsection{FLOW portrait construction}
FLOW portraits are constructed through several image processing steps that aim to extract ridges from an FTLE field (see Figure~\ref{fig:flow_ridge_extract} for a visualization of the intermediate processing steps).
It is important to note that we process the forward and backward FTLE fields separately and overlay them on the mean $\Delta F / F$ image to create the final FLOW portrait.

We begin by averaging the FTLE time-series to aggregate the flow features into mean forward and backward FTLE fields.
Next, we isolate possible ridge-like features by thresholding the mean FTLE field at a chosen percentile to form a binarized image.
This thresholding step is motivated by recognizing that a ridge can be though of as a continuous path along a local maximum in the field~\cite{shadden2005definition, LCSch3}.
Therefore, the binarized mean FTLE fields is thought to contain the ridges whose value is above the chosen threshold value.
Throughout this work we denote the specified threshold value as a parameter named the \emph{threshold percentile}.
For each FLOW portrait analysis we choose the threshold percentile to extract the FTLE ridges (see black arrows in Figure~\ref{fig:flow_ridge_extract}A for example ridges).
Figure~\ref{fig:flow_ridge_extract}A and B show the correspondence between the mean FTLE field and the binary versions (the threshold percentile was set to 95 percent).
We used threshold percentiles between $90 \%$ -- $93 \%$ for the mouse pup dataset and a threshold percentile of $93 \%$ for the adult mouse dataset.
Additionally, we thresholded the plane wave, the circular wave, and the traveling Gaussian examples to the $91$st, the $93$rd, and the $85$th percentiles respectively.
Supplemental Figure 4C illustrates FLOW portraits computes with a range of threshold percentiles.

Next, we perform two sets of morphological image processing operations on each binarized mean FTLE field to produce the final FLOW portrait.
The first set of operations aims denoise approximate ridges from the FTLE fields.
While the second set smooths the ridges to produce the FLOW portrait.
We found that these two series of operations provide strong approximations to the ridge features that we observe in the FTLE fields.
We use the \verb|bwmorph()| function in MATLAB for all morphological image processing operations(see \url{https://www.mathworks.com/help/images/ref/bwmorph.html} for details).
This function applies a specified morphological operation iteratively, with the number of iterations specified by the $n$ parameter, or until the input image remains unchanged, $n=$ Inf.
Unless otherwise specified, we performed morphological operations until the image no longer changed, with $n=$ Inf.
We refer the reader to the MATLAB documentation, Gonzales et al.~\cite{gonzalez2009Digital} and Haralick and Shapiro~\cite{haralick1992computer} for the mathematical details of each morphological processing operation used.

The first set of operations aims to transform the noisy, disconnected, ridges in the binarized images to connected ridges that reassemble those observed in the raw data.
First, we perform the \verb|`close'| operation (morphological dilation followed by erosion) to close any gaps within the binary image.
Next, we use the \verb|`thin'| operation to thin the blob-like structures seen in Figure~\ref{fig:flow_ridge_extract}B to a series of lines.
Lastly, we skeletonize the image by applying the \verb|`skel'| (performed with $n=4$).
Together these operations convert the disconnected, blob-like, structures seen in Figure~\ref{fig:flow_ridge_extract}B to the connected, single-pixel, structures in Figure~\ref{fig:flow_ridge_extract}C.
These skeletonized structures can be thought to approximate the centerlines of the FTLE ridges.

The second set of operations aims to smooth the skeletonized image to produce the FLOW portrait.
Here we performing the \verb|`diag'| operation to connect regions where two pixels lie corner-to-corner with an additional pixel.
We than apply the \verb|`spur'| operation to remove any remaining single pixel spurs from the ridges.
Lastly, we close any gaps introduced with the and the \verb|`close'| operation.

Lastly, we overlay the processed forward and backward images on the corresponding mean $\Delta F / F$ image to create the final FLOW portrait.
An example FLOW portrait can be seen in Figure~\ref{fig:flow_ridge_extract}D.

\subsection{Quantification of FLOW portrait consolidation during development}

In order to to quantify the consolidation of FLOW portraits during development we computed a metric we denote as the \emph{ridge count score}.
The ridge count score is defined as the number of disconnected ridges in the FLOW portrait divided by the total area of the FLOW portrait.
This metric is computed by counting disconnected ridges (objects) in the FLOW portrait and dividing by the total number of pixels included in the FLOW portrait.
This score was computed for forward FLOW, backward FLOW and the both combined for each sleep bout.
We then take the mean ridge count score of all sleep bouts from animals of the same developmental age.
Lastly, we use a paired t-test to determine whether the mean of mean ridge score for developmental days P1--P3 is statistically different than that for developmental days P5--P7.

\subsection{Code and data availability}

Our code is publicly available without restriction, other than citation, on Github at \url{https://github.com/natejlinden/FLOWPortrait}.
The code and data in this repository can reproduce all main analyses, findings, and figures from our paper.

\subsection*{Acknowledgements}
We are grateful for helpful discussion with Aditya Nair, Kameron Decker Harris, and Seth Hirsh.
NJL acknowledges support through a Neuroengineering Undergraduate Research Fellowship from the University of Washington Institute of Neuroengineering (UWIN) and the Washington Research Foundation Funds for Innovation in Neuroengineering.
DRT acknowledges funding support from UW Neuroscience Graduate Program (T32NS099578) and the UW Computational Neuroscience Center (5T90DA032436). NAS was supported by the Human Frontiers Science Program (Fellowship LT001071), and the European Union’s Horizon 2020 research and innovation programme (Marie Sklodowska-Curie
fellowship 656528).
WJM acknowledges funding from a Simons Foundation Autism grant.
SLB acknowledges funding support from the Army Research Office (ARO W911NF-19-1-0045).
BWB acknowledges funding from the Washington Research Foundation, the Alfred P. Sloan Foundation, and Weill Neurohub.

\subsection*{Author contributions}
NJL, SLB, and BWB conceived of the study and designed the analyses. NJL carried out the analyses. DRT, NAS, and WJM collected the imaging data and helped interpret the results. NJL and BWB wrote the paper, and all authors contributed to editing the manuscript.

\subsection*{Competing interests}
The authors have no competing interests.

 \begin{spacing}{.9}
 \small{
 \setlength{\bibsep}{6.5pt}
 \bibliographystyle{IEEEtran}
 \bibliography{article}

\begin{thebibliography}{10}
\providecommand{\url}[1]{#1}
\csname url@samestyle\endcsname
\providecommand{\newblock}{\relax}
\providecommand{\bibinfo}[2]{#2}
\providecommand{\BIBentrySTDinterwordspacing}{\spaceskip=0pt\relax}
\providecommand{\BIBentryALTinterwordstretchfactor}{4}
\providecommand{\BIBentryALTinterwordspacing}{\spaceskip=\fontdimen2\font plus
\BIBentryALTinterwordstretchfactor\fontdimen3\font minus
  \fontdimen4\font\relax}
\providecommand{\BIBforeignlanguage}[2]{{%
\expandafter\ifx\csname l@#1\endcsname\relax
\typeout{** WARNING: IEEEtran.bst: No hyphenation pattern has been}%
\typeout{** loaded for the language `#1'. Using the pattern for}%
\typeout{** the default language instead.}%
\else
\language=\csname l@#1\endcsname
\fi
#2}}
\providecommand{\BIBdecl}{\relax}
\BIBdecl

\bibitem{wekselblatt2016large}
J.~B. Wekselblatt, E.~D. Flister, D.~M. Piscopo, and C.~M. Niell, ``Large-scale
  imaging of cortical dynamics during sensory perception and behavior,''
  \emph{Journal of neurophysiology}, 2016.

\bibitem{muller2018cortical}
L.~Muller, F.~Chavane, J.~Reynolds, and T.~J. Sejnowski, ``Cortical travelling
  waves: mechanisms and computational principles,'' \emph{Nature Reviews
  Neuroscience}, vol.~19, no.~5, p. 255, 2018.

\bibitem{musall2019single}
S.~Musall, M.~T. Kaufman, A.~L. Juavinett, S.~Gluf, and A.~K. Churchland,
  ``Single-trial neural dynamics are dominated by richly varied movements,''
  \emph{Nature neuroscience}, vol.~22, no.~10, pp. 1677--1686, 2019.

\bibitem{corlew2004spontaneous}
R.~Corlew, M.~M. Bosma, and W.~J. Moody, ``Spontaneous, synchronous electrical
  activity in neonatal mouse cortical neurones,'' \emph{The Journal of
  physiology}, vol. 560, no.~2, pp. 377--390, 2004.

\bibitem{conhaim2010bimodal}
J.~Conhaim, E.~R. Cedarbaum, M.~Barahimi, J.~G. Moore, M.~I. Becker, H.~Gleiss,
  C.~Kohl, and W.~J. Moody, ``Bimodal septal and cortical triggering and
  complex propagation patterns of spontaneous waves of activity in the
  developing mouse cerebral cortex,'' \emph{Developmental neurobiology},
  vol.~70, no.~10, pp. 679--692, 2010.

\bibitem{luhmann2016spontaneous}
H.~J. Luhmann, A.~Sinning, J.-W. Yang, V.~Reyes-Puerta, M.~C. St{\"u}ttgen,
  S.~Kirischuk, and W.~Kilb, ``Spontaneous neuronal activity in developing
  neocortical networks: from single cells to large-scale interactions,''
  \emph{Frontiers in neural circuits}, vol.~10, p.~40, 2016.

\bibitem{rossi2017focal}
L.~F. Rossi, R.~C. Wykes, D.~M. Kullmann, and M.~Carandini, ``Focal cortical
  seizures start as standing waves and propagate respecting homotopic
  connectivity,'' \emph{Nature communications}, vol.~8, no.~1, pp. 1--11, 2017.

\bibitem{cramer2019vivo}
J.~V. Cramer, B.~Gesierich, S.~Roth, M.~Dichgans, M.~D{\"u}ring, and A.~Liesz,
  ``In vivo widefield calcium imaging of the mouse cortex for analysis of
  network connectivity in health and brain disease,'' \emph{Neuroimage}, vol.
  199, pp. 570--584, 2019.

\bibitem{mcgirr2017cortical}
A.~McGirr, J.~LeDue, A.~W. Chan, Y.~Xie, and T.~H. Murphy, ``Cortical
  functional hyperconnectivity in a mouse model of depression and selective
  network effects of ketamine,'' \emph{Brain}, vol. 140, no.~8, pp. 2210--2225,
  2017.

\bibitem{siapas1998coordinated}
A.~G. Siapas and M.~A. Wilson, ``Coordinated interactions between hippocampal
  ripples and cortical spindles during slow-wave sleep,'' \emph{Neuron},
  vol.~21, no.~5, pp. 1123--1128, 1998.

\bibitem{wekselblatt2019distinct}
J.~B. Wekselblatt and C.~M. Niell, ``Distinct functional classes of excitatory
  neurons in mouse {V}1 are differentially modulated by learning and task
  engagement,'' \emph{bioRxiv}, p. 533463, 2019.

\bibitem{liu2019assessing}
M.~Liu, C.~Song, Y.~Liang, T.~Kn{\"o}pfel, and C.~Zhou, ``Assessing
  spatiotemporal variability of brain spontaneous activity by multiscale
  entropy and functional connectivity,'' \emph{NeuroImage}, vol. 198, pp.
  198--220, 2019.

\bibitem{Feller1996requirement}
M.~B. Feller, D.~P. Wellis, D.~Stellwagen, F.~S. Werblin, and C.~J. Shatz,
  ``Requirement for cholinergic synaptic transmission in the propagation of
  spontaneous retinal waves,'' \emph{Science}, vol. 272, no. 5265, pp.
  1182--1187, 1996.

\bibitem{feller1997dynamic}
M.~B. Feller, D.~A. Butts, H.~L. Aaron, D.~S. Rokhsar, and C.~J. Shatz,
  ``Dynamic processes shape spatiotemporal properties of retinal waves,''
  \emph{Neuron}, vol.~19, no.~2, pp. 293--306, 1997.

\bibitem{wong1999retinal}
R.~O. Wong, ``Retinal waves and visual system development,'' \emph{Annual
  review of neuroscience}, vol.~22, no.~1, pp. 29--47, 1999.

\bibitem{tiriac2018light}
A.~Tiriac, B.~E. Smith, and M.~B. Feller, ``Light prior to eye opening promotes
  retinal waves and eye-specific segregation,'' \emph{Neuron}, vol. 100, no.~5,
  pp. 1059--1065, 2018.

\bibitem{ren2021characterizing}
C.~Ren and T.~Komiyama, ``Characterizing cortex-wide dynamics with wide-field
  calcium imaging,'' \emph{Journal of Neuroscience}, vol.~41, no.~19, pp.
  4160--4168, 2021.

\bibitem{urai2021large}
A.~E. Urai, B.~Doiron, A.~M. Leifer, and A.~K. Churchland, ``Large-scale neural
  recordings call for new insights to link brain and behavior,'' \emph{arXiv
  preprint arXiv:2103.14662}, 2021.

\bibitem{sato2012traveling}
T.~K. Sato, I.~Nauhaus, and M.~Carandini, ``Traveling waves in visual cortex,''
  \emph{Neuron}, vol.~75, no.~2, pp. 218--229, 2012.

\bibitem{dana2014thy1}
H.~Dana, T.-W. Chen, A.~Hu, B.~C. Shields, C.~Guo, L.~L. Looger, D.~S. Kim, and
  K.~Svoboda, ``Thy1-{GC}a{MP}6 transgenic mice for neuronal population imaging
  in vivo,'' \emph{PloS one}, vol.~9, no.~9, 2014.

\bibitem{stirman2016wide}
J.~N. Stirman, I.~T. Smith, M.~W. Kudenov, and S.~L. Smith, ``Wide
  field-of-view, multi-region, two-photon imaging of neuronal activity in the
  mammalian brain,'' \emph{Nature biotechnology}, vol.~34, no.~8, pp. 857--862,
  2016.

\bibitem{silasi2016intact}
G.~Silasi, D.~Xiao, M.~P. Vanni, A.~C. Chen, and T.~H. Murphy, ``Intact skull
  chronic windows for mesoscopic wide-field imaging in awake mice,''
  \emph{Journal of neuroscience methods}, vol. 267, pp. 141--149, 2016.

\bibitem{steinmetz2017aberrant}
N.~A. Steinmetz, C.~Buetfering, J.~Lecoq, C.~R. Lee, A.~J. Peters, E.~A.
  Jacobs, P.~Coen, D.~R. Ollerenshaw, M.~T. Valley, S.~E. De~Vries
  \emph{et~al.}, ``Aberrant cortical activity in multiple {GC}a{MP}6-expressing
  transgenic mouse lines,'' \emph{eneuro}, 2017.

\bibitem{couto2021chronic}
J.~Couto, S.~Musall, X.~R. Sun, A.~Khanal, S.~Gluf, S.~Saxena, I.~Kinsella,
  T.~Abe, J.~P. Cunningham, L.~Paninski \emph{et~al.}, ``Chronic, cortex-wide
  imaging of specific cell populations during behavior,'' \emph{Nature
  Protocols}, pp. 1--25, 2021.

\bibitem{nakai2001high}
J.~Nakai, M.~Ohkura, and K.~Imoto, ``A high signal-to-noise {C}a 2+ probe
  composed of a single green fluorescent protein,'' \emph{Nature
  biotechnology}, vol.~19, no.~2, pp. 137--141, 2001.

\bibitem{tian2009imaging}
L.~Tian, S.~A. Hires, T.~Mao, D.~Huber, M.~E. Chiappe, S.~H. Chalasani,
  L.~Petreanu, J.~Akerboom, S.~A. McKinney, E.~R. Schreiter, C.~I. Bargmann,
  V.~Jayaraman, K.~Svoboda, and L.~L. Looger, ``Imaging neural activity in
  worms, flies and mice with improved {GC}a{MP} calcium indicators,''
  \emph{Nature methods}, vol.~6, no.~12, p. 875, 2009.

\bibitem{chen2013ultrasensitive}
T.-W. Chen, T.~J. Wardill, Y.~Sun, S.~R. Pulver, S.~L. Renninger, A.~Baohan,
  E.~R. Schreiter, R.~A. Kerr, M.~B. Orger, V.~Jayaraman, L.~L. Looger,
  K.~Svoboda, and D.~S. Kim, ``Ultrasensitive fluorescent proteins for imaging
  neuronal activity,'' \emph{Nature}, vol. 499, no. 7458, pp. 295--300, 2013.

\bibitem{vanni2014mesoscale}
M.~P. Vanni and T.~H. Murphy, ``Mesoscale transcranial spontaneous activity
  mapping in {GC}a{MP}3 transgenic mice reveals extensive reciprocal
  connections between areas of somatomotor cortex,'' \emph{Journal of
  Neuroscience}, vol.~34, no.~48, pp. 15\,931--15\,946, 2014.

\bibitem{mcvea2012voltage}
D.~A. McVea, M.~H. Mohajerani, and T.~H. Murphy, ``Voltage-sensitive dye
  imaging reveals dynamic spatiotemporal properties of cortical activity after
  spontaneous muscle twitches in the newborn rat,'' \emph{Journal of
  Neuroscience}, vol.~32, no.~32, pp. 10\,982--10\,994, 2012.

\bibitem{song2018cortical}
C.~Song, D.~M. Piscopo, C.~M. Niell, and T.~Kn{\"o}pfel, ``Cortical signatures
  of wakeful somatosensory processing,'' \emph{Scientific reports}, vol.~8,
  no.~1, pp. 1--12, 2018.

\bibitem{scott2018imaging}
B.~B. Scott, S.~Y. Thiberge, C.~Guo, D.~G.~R. Tervo, C.~D. Brody, A.~Y.
  Karpova, and D.~W. Tank, ``Imaging cortical dynamics in {GC}a{MP} transgenic
  rats with a head-mounted widefield macroscope,'' \emph{Neuron}, vol. 100,
  no.~5, pp. 1045--1058, 2018.

\bibitem{allen2017global}
W.~E. Allen, I.~V. Kauvar, M.~Z. Chen, E.~B. Richman, S.~J. Yang, K.~Chan,
  V.~Gradinaru, B.~E. Deverman, L.~Luo, and K.~Deisseroth, ``Global
  representations of goal-directed behavior in distinct cell types of mouse
  neocortex,'' \emph{Neuron}, vol.~94, no.~4, pp. 891--907, 2017.

\bibitem{pinto2019task}
L.~Pinto, K.~Rajan, B.~DePasquale, S.~Y. Thiberge, D.~W. Tank, and C.~D. Brody,
  ``Task-dependent changes in the large-scale dynamics and necessity of
  cortical regions,'' \emph{Neuron}, vol. 104, no.~4, pp. 810--824, 2019.

\bibitem{jacobs2018cortical}
E.~A. Jacobs, N.~A. Steinmetz, M.~Carandini, and K.~D. Harris, ``Cortical state
  fluctuations during sensory decision making,'' \emph{Biorxiv}, p. 348193,
  2018.

\bibitem{zatka2020perceptual}
P.~Zatka-Haas, N.~A. Steinmetz, M.~Carandini, and K.~D. Harris, ``A perceptual
  decision requires sensory but not action coding in mouse cortex,''
  \emph{bioRxiv}, p. 501627, 2020.

\bibitem{wright2017functional}
P.~W. Wright, L.~M. Brier, A.~Q. Bauer, G.~A. Baxter, A.~W. Kraft, M.~D.
  Reisman, A.~R. Bice, A.~Z. Snyder, J.-M. Lee, and J.~P. Culver, ``Functional
  connectivity structure of cortical calcium dynamics in anesthetized and awake
  mice,'' \emph{PloS one}, vol.~12, no.~10, 2017.

\bibitem{vanni2017mesoscale}
M.~P. Vanni, A.~W. Chan, M.~Balbi, G.~Silasi, and T.~H. Murphy, ``Mesoscale
  mapping of mouse cortex reveals frequency-dependent cycling between distinct
  macroscale functional modules,'' \emph{Journal of Neuroscience}, vol.~37,
  no.~31, pp. 7513--7533, 2017.

\bibitem{tabuena2019manuscript}
D.~R. Tabuena, R.~Huynh, J.~Metcalf, T.~Richner, A.~Stroh, B.~W. Brunton, W.~J.
  Moody, and C.~R. Easton, ``Pan-cortical waves in the neonatal rodent brain
  \emph{in vivo}: A precursor of adult sleep waves?'' \emph{\emph{In review.}},
  2019.

\bibitem{guckenheimer_holmes}
P.~Holmes and J.~Guckenheimer, \emph{Nonlinear oscillations, dynamical systems,
  and bifurcations of vector fields}, ser. Applied Mathematical Sciences.\hskip
  1em plus 0.5em minus 0.4em\relax Berlin, Heidelberg: Springer-Verlag, 1983,
  vol.~42.

\bibitem{pang2016dimensionality}
R.~Pang, B.~J. Lansdell, and A.~L. Fairhall, ``Dimensionality reduction in
  neuroscience,'' \emph{Current Biology}, vol.~26, no.~14, pp. R656--R660,
  2016.

\bibitem{cunningham2014dimensionality}
J.~P. Cunningham and M.~Y. Byron, ``Dimensionality reduction for large-scale
  neural recordings,'' \emph{Nature neuroscience}, vol.~17, no.~11, pp.
  1500--1509, 2014.

\bibitem{dyer2017cryptography}
E.~L. Dyer, M.~G. Azar, M.~G. Perich, H.~L. Fernandes, S.~Naufel, L.~E. Miller,
  and K.~P. K{\"o}rding, ``A cryptography-based approach for movement
  decoding,'' \emph{Nature Biomedical Engineering}, vol.~1, no.~12, pp.
  967--976, 2017.

\bibitem{ganguli2012compressed}
S.~Ganguli and H.~Sompolinsky, ``Compressed sensing, sparsity, and
  dimensionality in neuronal information processing and data analysis,''
  \emph{Annual review of neuroscience}, vol.~35, pp. 485--508, 2012.

\bibitem{churchland2012neural}
M.~M. Churchland, J.~P. Cunningham, M.~T. Kaufman, J.~D. Foster, P.~Nuyujukian,
  S.~I. Ryu, and K.~V. Shenoy, ``Neural population dynamics during reaching,''
  \emph{Nature}, vol. 487, no. 7405, pp. 51--56, 2012.

\bibitem{gallego2017neural}
J.~A. Gallego, M.~G. Perich, L.~E. Miller, and S.~A. Solla, ``Neural manifolds
  for the control of movement,'' \emph{Neuron}, vol.~94, no.~5, pp. 978--984,
  2017.

\bibitem{gallego2020long}
J.~A. Gallego, M.~G. Perich, R.~H. Chowdhury, S.~A. Solla, and L.~E. Miller,
  ``Long-term stability of cortical population dynamics underlying consistent
  behavior,'' \emph{Nature neuroscience}, vol.~23, no.~2, pp. 260--270, 2020.

\bibitem{cunningham2015linear}
J.~P. Cunningham and Z.~Ghahramani, ``Linear dimensionality reduction: Survey,
  insights, and generalizations,'' \emph{The Journal of Machine Learning
  Research}, vol.~16, no.~1, pp. 2859--2900, 2015.

\bibitem{brunton2016extracting}
B.~W. Brunton, L.~A. Johnson, J.~G. Ojemann, and J.~N. Kutz, ``Extracting
  spatial--temporal coherent patterns in large-scale neural recordings using
  dynamic mode decomposition,'' \emph{Journal of neuroscience methods}, vol.
  258, pp. 1--15, 2016.

\bibitem{Tu2014jcd}
J.~H. Tu, C.~W. Rowley, D.~M. Luchtenburg, S.~L. Brunton, and J.~N. Kutz, ``On
  dynamic mode decomposition: theory and applications,'' \emph{Journal of
  Computational Dynamics}, vol.~1, no.~2, pp. 391--421, 2014.

\bibitem{kutz2016dynamic}
J.~N. Kutz, S.~L. Brunton, B.~W. Brunton, and J.~L. Proctor, \emph{Dynamic mode
  decomposition: data-driven modeling of complex systems}.\hskip 1em plus 0.5em
  minus 0.4em\relax SIAM, 2016.

\bibitem{macdowell2020low}
C.~J. MacDowell and T.~J. Buschman, ``Low-dimensional spatiotemporal dynamics
  underlie cortex-wide neural activity,'' \emph{Current Biology}, 2020.

\bibitem{saxena2020localized}
S.~Saxena, I.~Kinsella, S.~Musall, S.~H. Kim, J.~Meszaros, D.~N. Thibodeaux,
  C.~Kim, J.~Cunningham, E.~M. Hillman, A.~Churchland \emph{et~al.},
  ``Localized semi-nonnegative matrix factorization ({L}oca{NMF}) of widefield
  calcium imaging data,'' \emph{PLOS Computational Biology}, vol.~16, no.~4, p.
  e1007791, 2020.

\bibitem{mackevicius2019unsupervised}
E.~L. Mackevicius, A.~H. Bahle, A.~H. Williams, S.~Gu, N.~I. Denisenko, M.~S.
  Goldman, and M.~S. Fee, ``Unsupervised discovery of temporal sequences in
  high-dimensional datasets, with applications to neuroscience,'' \emph{Elife},
  vol.~8, p. e38471, 2019.

\bibitem{zhou2018efficient}
P.~Zhou, S.~L. Resendez, J.~Rodriguez-Romaguera, J.~C. Jimenez, S.~Q. Neufeld,
  A.~Giovannucci, J.~Friedrich, E.~A. Pnevmatikakis, G.~D. Stuber, R.~Hen
  \emph{et~al.}, ``Efficient and accurate extraction of in vivo calcium signals
  from microendoscopic video data,'' \emph{Elife}, vol.~7, p. e28728, 2018.

\bibitem{zanos2015sensorimotor}
T.~P. Zanos, P.~J. Mineault, K.~T. Nasiotis, D.~Guitton, and C.~C. Pack, ``A
  sensorimotor role for traveling waves in primate visual cortex,''
  \emph{Neuron}, vol.~85, no.~3, pp. 615--627, 2015.

\bibitem{blankenship2011role}
A.~G. Blankenship, A.~M. Hamby, A.~Firl, S.~Vyas, S.~Maxeiner, K.~Willecke, and
  M.~B. Feller, ``The role of neuronal connexins 36 and 45 in shaping
  spontaneous firing patterns in the developing retina,'' \emph{Journal of
  Neuroscience}, vol.~31, no.~27, pp. 9998--10\,008, 2011.

\bibitem{afrashteh2017optical}
N.~Afrashteh, S.~Inayat, M.~Mohsenvand, and M.~H. Mohajerani, ``Optical-flow
  analysis toolbox for characterization of spatiotemporal dynamics in mesoscale
  optical imaging of brain activity,'' \emph{Neuroimage}, vol. 153, pp. 58--74,
  2017.

\bibitem{townsend2018detection}
R.~G. Townsend and P.~Gong, ``Detection and analysis of spatiotemporal patterns
  in brain activity,'' \emph{PLoS computational biology}, vol.~14, no.~12, p.
  e1006643, 2018.

\bibitem{Haller2002pof}
G.~Haller, ``{Lagrangian} coherent structures from approximate velocity data,''
  \emph{Physics of Fluids}, vol.~14, no.~6, pp. 1851--1861, June 2002.

\bibitem{LCSch3}
\BIBentryALTinterwordspacing
S.~C. Shadden, \emph{Lagrangian Coherent Structures}.\hskip 1em plus 0.5em
  minus 0.4em\relax John Wiley \& Sons, Ltd, 2011, ch.~3, pp. 59--89. [Online].
  Available:
  \url{https://onlinelibrary.wiley.com/doi/abs/10.1002/9783527639748.ch3}
\BIBentrySTDinterwordspacing

\bibitem{haller2015lagrangian}
G.~Haller, ``Lagrangian coherent structures,'' \emph{Annual Review of Fluid
  Mechanics}, vol.~47, pp. 137--162, 2015.

\bibitem{shadden2005definition}
S.~C. Shadden, F.~Lekien, and J.~E. Marsden, ``Definition and properties of
  lagrangian coherent structures from finite-time lyapunov exponents in
  two-dimensional aperiodic flows,'' \emph{Physica D: Nonlinear Phenomena},
  vol. 212, no. 3-4, pp. 271--304, 2005.

\bibitem{Mathur2007prl}
M.~Mathur, G.~Haller, T.~Peacock, J.~E. Ruppert-Felsot, and H.~L. Swinney,
  ``Uncovering the {Lagrangian} skeleton of turbulence,'' \emph{Physical Review
  Letters}, vol.~98, pp. 144\,502--1--144\,502--4, 2007.

\bibitem{Green2007jfm}
M.~A. Green, C.~W. Rowley, and G.~Haller, ``Detection of {Lagrangian} coherent
  structures in {3D} turbulence.'' \emph{Journal of Fluid Mechanics}, vol. 572,
  pp. 111--120, 2007.

\bibitem{Brunton2010chaos}
S.~L. Brunton and C.~W. Rowley, ``Fast computation of {FTLE} fields for
  unsteady flows: a comparison of methods,'' \emph{Chaos}, vol.~20, p. 017503,
  2010.

\bibitem{Farazmand2012chaos}
M.~Farazmand and G.~Haller, ``Computing {L}agrangian coherent structures from
  their variational theory,'' \emph{Chaos}, vol.~22, no. 013128, pp.
  013\,128--1--013\,128--12, 2012.

\bibitem{peng2009transport}
J.~Peng and J.~Dabiri, ``Transport of inertial particles by lagrangian coherent
  structures: application to predator--prey interaction in jellyfish feeding,''
  \emph{Journal of Fluid Mechanics}, vol. 623, pp. 75--84, 2009.

\bibitem{duvernois2013lagrangian}
V.~Duvernois, A.~L. Marsden, and S.~C. Shadden, ``Lagrangian analysis of
  hemodynamics data from fsi simulation,'' \emph{International journal for
  numerical methods in biomedical engineering}, vol.~29, no.~4, pp. 445--461,
  2013.

\bibitem{shadden2015lagrangian}
S.~C. Shadden and A.~Arzani, ``Lagrangian postprocessing of computational
  hemodynamics,'' \emph{Annals of biomedical engineering}, vol.~43, no.~1, pp.
  41--58, 2015.

\bibitem{mohajerani2013spontaneous}
M.~H. Mohajerani, A.~W. Chan, M.~Mohsenvand, J.~LeDue, R.~Liu, D.~A. McVea,
  J.~D. Boyd, Y.~T. Wang, M.~Reimers, and T.~H. Murphy, ``Spontaneous cortical
  activity alternates between motifs defined by regional axonal projections,''
  \emph{Nature neuroscience}, vol.~16, no.~10, p. 1426, 2013.

\bibitem{ashby2019peripheral}
D.~M. Ashby, J.~LeDue, T.~H. Murphy, and A.~McGirr, ``Peripheral nerve ligation
  elicits widespread alterations in cortical sensory evoked and spontaneous
  activity,'' \emph{Scientific reports}, vol.~9, no.~1, pp. 1--10, 2019.

\bibitem{jia2011vivo}
H.~Jia, N.~L. Rochefort, X.~Chen, and A.~Konnerth, ``In vivo two-photon imaging
  of sensory-evoked dendritic calcium signals in cortical neurons,''
  \emph{Nature protocols}, vol.~6, no.~1, pp. 28--35, 2011.

\bibitem{onu2015lcs}
K.~Onu, F.~Huhn, and G.~Haller, ``{LCS Tool:} a computational platform for
  lagrangian coherent structures,'' \emph{Journal of Computational Science},
  vol.~7, pp. 26--36, 2015.

\bibitem{paragios2006handbook}
N.~Paragios, Y.~Chen, and O.~D. Faugeras, \emph{Handbook of mathematical models
  in computer vision}.\hskip 1em plus 0.5em minus 0.4em\relax Springer Science
  \& Business Media, 2006.

\bibitem{duffy1991sensitivity}
C.~J. Duffy and R.~H. Wurtz, ``Sensitivity of {MST} neurons to optic flow
  stimuli. {II. Mechanisms of} response selectivity revealed by small-field
  stimuli,'' \emph{Journal of neurophysiology}, vol.~65, no.~6, pp. 1346--1359,
  1991.

\bibitem{krapp1996estimation}
H.~G. Krapp and R.~Hengstenberg, ``Estimation of self-motion by optic flow
  processing in single visual interneurons,'' \emph{Nature}, vol. 384, no.
  6608, pp. 463--466, 1996.

\bibitem{horn1981determining}
B.~K. Horn and B.~G. Schunck, ``Determining optical flow,'' in \emph{Techniques
  and Applications of Image Understanding}, vol. 281.\hskip 1em plus 0.5em
  minus 0.4em\relax International Society for Optics and Photonics, 1981, pp.
  319--331.

\bibitem{willert1991digital}
C.~E. Willert and M.~Gharib, ``Digital particle image velocimetry,''
  \emph{Experiments in fluids}, vol.~10, no.~4, pp. 181--193, 1991.

\bibitem{westerweel1997fundamentals}
J.~Westerweel, ``Fundamentals of digital particle image velocimetry,''
  \emph{Measurement science and technology}, vol.~8, no.~12, p. 1379, 1997.

\bibitem{Taira2017aiaa}
K.~Taira, S.~L. Brunton, S.~Dawson, C.~W. Rowley, T.~Colonius, B.~J. McKeon,
  O.~T. Schmidt, S.~Gordeyev, V.~Theofilis, and L.~S. Ukeiley, ``Modal analysis
  of fluid flows: An overview,'' \emph{AIAA Journal}, vol.~55, no.~12, pp.
  4013--4041, 2017.

\bibitem{Taira2020aiaa}
K.~Taira, M.~S. Hemati, S.~L. Brunton, Y.~Sun, K.~Duraisamy, S.~Bagheri,
  S.~Dawson, and C.-A. Yeh, ``Modal analysis of fluid flows: Applications and
  outlook,'' \emph{AIAA Journal}, vol.~58, no.~3, pp. 998--1022, 2020.

\bibitem{garth2007efficient}
C.~Garth, F.~Gerhardt, X.~Tricoche, and H.~Hans, ``Efficient computation and
  visualization of coherent structures in fluid flow applications,'' \emph{IEEE
  Transactions on Visualization and Computer Graphics}, vol.~13, no.~6, pp.
  1464--1471, 2007.

\bibitem{lipinski2010ridge}
D.~Lipinski and K.~Mohseni, ``A ridge tracking algorithm and error estimate for
  efficient computation of lagrangian coherent structures,'' \emph{Chaos: An
  Interdisciplinary Journal of Nonlinear Science}, vol.~20, no.~1, p. 017504,
  2010.

\bibitem{garaschuk2000large}
O.~Garaschuk, J.~Linn, J.~Eilers, and A.~Konnerth, ``Large-scale oscillatory
  calcium waves in the immature cortex,'' \emph{Nature neuroscience}, vol.~3,
  no.~5, pp. 452--459, 2000.

\bibitem{conhaim2011developmental}
J.~Conhaim, C.~R. Easton, M.~I. Becker, M.~Barahimi, E.~R. Cedarbaum, J.~G.
  Moore, L.~F. Mather, S.~Dabagh, D.~J. Minter, S.~P. Moen \emph{et~al.},
  ``Developmental changes in propagation patterns and transmitter dependence of
  waves of spontaneous activity in the mouse cerebral cortex,'' \emph{The
  Journal of physiology}, vol. 589, no.~10, pp. 2529--2541, 2011.

\bibitem{easton2014genetic}
C.~R. Easton, K.~Weir, A.~Scott, S.~P. Moen, Z.~Barger, A.~Folch, R.~F. Hevner,
  and W.~J. Moody, ``Genetic elimination of gabaergic neurotransmission reveals
  two distinct pacemakers for spontaneous waves of activity in the developing
  mouse cortex,'' \emph{Journal of Neuroscience}, vol.~34, no.~11, pp.
  3854--3863, 2014.

\bibitem{barger2016early}
Z.~Barger, C.~R. Easton, K.~E. Neuzil, and W.~J. Moody, ``Early network
  activity propagates bidirectionally between hippocampus and cortex,''
  \emph{Developmental neurobiology}, vol.~76, no.~6, pp. 661--672, 2016.

\bibitem{seelke2010developmental}
A.~M. Seelke and M.~S. Blumberg, ``Developmental appearance and disappearance
  of cortical events and oscillations in infant rats,'' \emph{Brain research},
  vol. 1324, pp. 34--42, 2010.

\bibitem{blumberg2014development}
M.~S. Blumberg, A.~J. Gall, and W.~D. Todd, ``The development of sleep--wake
  rhythms and the search for elemental circuits in the infant brain.''
  \emph{Behavioral neuroscience}, vol. 128, no.~3, p. 250, 2014.

\bibitem{wang2020allen}
Q.~Wang, S.-L. Ding, Y.~Li, J.~Royall, D.~Feng, P.~Lesnar, N.~Graddis,
  M.~Naeemi, B.~Facer, A.~Ho \emph{et~al.}, ``{The Allen Mouse Brain Common
  Coordinate Framework: A 3D} reference atlas,'' \emph{Cell}, 2020.

\bibitem{rudy2017data}
S.~H. Rudy, S.~L. Brunton, J.~L. Proctor, and J.~N. Kutz, ``Data-driven
  discovery of partial differential equations,'' \emph{Science Advances},
  vol.~3, no.~4, p. e1602614, 2017.

\bibitem{Schaeffer2017prsa}
H.~Schaeffer, ``Learning partial differential equations via data discovery and
  sparse optimization,'' in \emph{Proc. R. Soc. A}, vol. 473, no. 2197.\hskip
  1em plus 0.5em minus 0.4em\relax The Royal Society, 2017, p. 20160446.

\bibitem{hsofcode}
\BIBentryALTinterwordspacing
M.~Kharbat, ``{Horn-Schunck} optical flow method,'' \emph{MATLAB Central File
  Exchange}, 2009. [Online]. Available:
  \url{https://www.mathworks.com/matlabcentral/fileexchange/22756-horn-schunck-optical-flow-method}
\BIBentrySTDinterwordspacing

\bibitem{gonzalez2009Digital}
C.~Gonzalez, R, E.~Woods, R, and L.~Eddins~S, \emph{Digital Image Processing
  Using MATLAB}.\hskip 1em plus 0.5em minus 0.4em\relax Gatesmark Publishing,
  2009.

\bibitem{haralick1992computer}
M.~Haralick, Robert and G.~Shapiro, Linda, \emph{Computer and Robot Vision, Vol
  1}.\hskip 1em plus 0.5em minus 0.4em\relax Addison-Wesley, 1992.

\end{thebibliography}


\begin{thebibliography}{1}
\providecommand{\url}[1]{#1}
\csname url@samestyle\endcsname
\providecommand{\newblock}{\relax}
\providecommand{\bibinfo}[2]{#2}
\providecommand{\BIBentrySTDinterwordspacing}{\spaceskip=0pt\relax}
\providecommand{\BIBentryALTinterwordstretchfactor}{4}
\providecommand{\BIBentryALTinterwordspacing}{\spaceskip=\fontdimen2\font plus
\BIBentryALTinterwordstretchfactor\fontdimen3\font minus
  \fontdimen4\font\relax}
\providecommand{\BIBforeignlanguage}[2]{{%
\expandafter\ifx\csname l@#1\endcsname\relax
\typeout{** WARNING: IEEEtran.bst: No hyphenation pattern has been}%
\typeout{** loaded for the language `#1'. Using the pattern for}%
\typeout{** the default language instead.}%
\else
\language=\csname l@#1\endcsname
\fi
#2}}
\providecommand{\BIBdecl}{\relax}
\BIBdecl

\bibitem{tabuena2019manuscript}
D.~R. Tabuena, R.~Huynh, J.~Metcalf, T.~Richner, A.~Stroh, B.~W. Brunton, W.~J.
  Moody, and C.~R. Easton, ``Pan-cortical waves in the neonatal rodent brain
  \emph{in vivo}: A precursor of adult sleep waves?'' \emph{\emph{In review.}},
  2019.

\bibitem{zatka2020perceptual}
P.~Zatka-Haas, N.~A. Steinmetz, M.~Carandini, and K.~D. Harris, ``A perceptual
  decision requires sensory but not action coding in mouse cortex,''
  \emph{bioRxiv}, p. 501627, 2020.

\bibitem{afrashteh2017optical}
N.~Afrashteh, S.~Inayat, M.~Mohsenvand, and M.~H. Mohajerani, ``Optical-flow
  analysis toolbox for characterization of spatiotemporal dynamics in mesoscale
  optical imaging of brain activity,'' \emph{Neuroimage}, vol. 153, pp. 58--74,
  2017.

\end{thebibliography}
 }
 \end{spacing}
\end{document}


\maketitle

\blfootnote{$^*$ Corresponding author: \url{bbrunton@uw.edu}.}

\section{Code and Data Access}

All code and data developed to construct FLOW portraits and to run analyses shown here is freely available in a repository at \url{https://github.com/natejlinden/FLOWPortrait}.
The code is implemented in MATLAB 2017b.
A basic MATLAB implementation with  MATLAB's Image Processing Toolbox (\url{https://www.mathworks.com/products/image.html}) is enough to reproduce the figures and run additional FLOW Portrait analyses.
The data required to created the figures from the developing mouse dataset~\cite{tabuena2019manuscript} and from the adult mouse dataset~\cite{zatka2020perceptual} are included in the repository.

\section{Supplemental Videos}

Supplemental vidoes 1--5 are available in the repository at \url{https://github.com/natejlinden/FLOWPortrait/tree/master/SupplementaryVideos}.

\subsubsection*{Video captions:}
\begin{itemize}
    \item Video 1 -- Steps to compute a FLOW portrait for a synthetic two-dimensional Gaussian that grows, translates to the right, then shrinks. This synthetic example is depicted in Figure 1, and Supplemental Figures 2 and 4.  (\textbf{Raw Data}) Raw spatiotemporal data for the two-dimensional Gaussian example.  (\textbf{Optical Flow}) The optical flow vector field overlaid on the original data. Optical flow is used to convert frame-by-frame changes in pixel intensity into a vector field. (\textbf{Particles}) Uniformly spaced particles are propagated through space according to the optical flow vector field. The particles' positions are updated by adding a displacement that corresponds to the vector field multiplied by a unit time step. Regions where particles cluster will be captured as ridges in the backward time FTLE and regions where particles separate will be captured as ridges in the forward time FTLE. (\textbf{FLOW Portrait}) The final FLOW portrait for this dataset. Orange structures correspond to the forward time FTLE and show regions where activity is repelled; purple structures correspond to the backward time FTLE and show regions where activity is attracted. The FLOW portrait was computed using a XX frame integration length and a XXth threshold percentile.
    \item Video 2 -- An example pan cortical wave (left pane) in widefield imaging data recorded from a developing mouse and the corresponding FLOW portrait (right pane). This wave event is depicted in Figure 6B as wave ii. The video is played at 20 frames per second such that one second of video corresponds to one second of recording. The FLOW portrait was computed using an integration length of 2 seconds (40 frames), and a visualization threshold of 0.92.
    \item Video 3 -- An example pan cortical wave (left pane) in widefield imaging data recorded from a developing mouse and the corresponding FLOW portrait (right pane). This wave event is depicted in Figure 6B as wave i. Note, the FLOW portrait differs from that seen in the figure due to differences in thresholding. The video is played at 20 frames per second such that one second of video corresponds to one second of recording. The FLOW portrait was computed using an integration length of 2 seconds (40 frames), and a visualization threshold of 0.92.
    \item Video 4 -- An example bout of spontaneous (left pane) cortical activity in in widefield imaging data recorded from an adult mouse and the corresponding FLOW portrait (right pane). This bout is depicted in Figure 8 as bout i. The video is played at 20 frames per second such that one second of video corresponds to one second of recording. The FLOW portrait was computed using an integration length of approximately 0.4 seconds (15 frames), and a visualization threshold of 0.93.
    \item Video 5 -- An example bout of spontaneous (left pane) cortical activity in in widefield imaging data recorded from an adult mouse and the corresponding FLOW portrait (right pane). This bout is depicted in Figure b as bout ii. The video is played at 20 frames per second such that one second of video corresponds to one second of recording. The FLOW portrait was computed using an integration length of approximately 0.4 seconds (15 frames), and a visualization threshold of 0.93.
\end{itemize}

\clearpage
\section{Supplemental Figures}
\begin{figure}[htp]
    \centering
    \includegraphics[width=\textwidth]{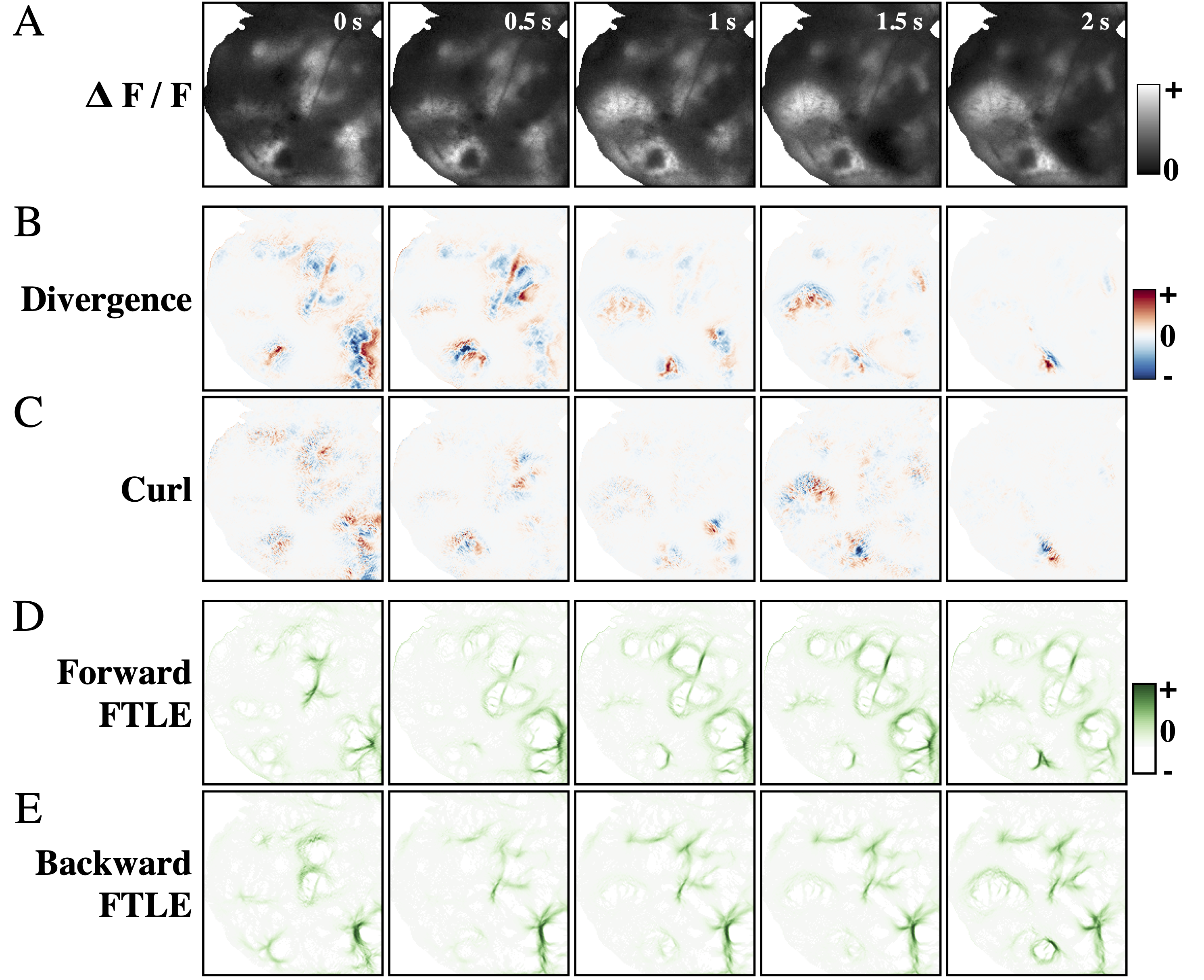}
    \caption{The unsteady nature of widefield imaging data suggested Lagrangian metrics (FTLE) will prove more useful than instantaneous metrics (divergence and curl) in summarizing activity.
    (\textbf{A}) Widefield imaging data from which instantaneous divergence and curl along with the Lagrangian FTLE are computed. Instantaneous divergence (\textbf{B}) and curl (\textbf{C}) show a similar unsteady nature to the data.
    Meanwhile, persistent features in the forward and backward FTLEs (\textbf{D} and \textbf{E}, respectively) can summarize the activity seen throughout the segment of data.}
    \label{vector_calc}
\end{figure}

\begin{figure}[htp]
    \centering
    \includegraphics[width=\textwidth]{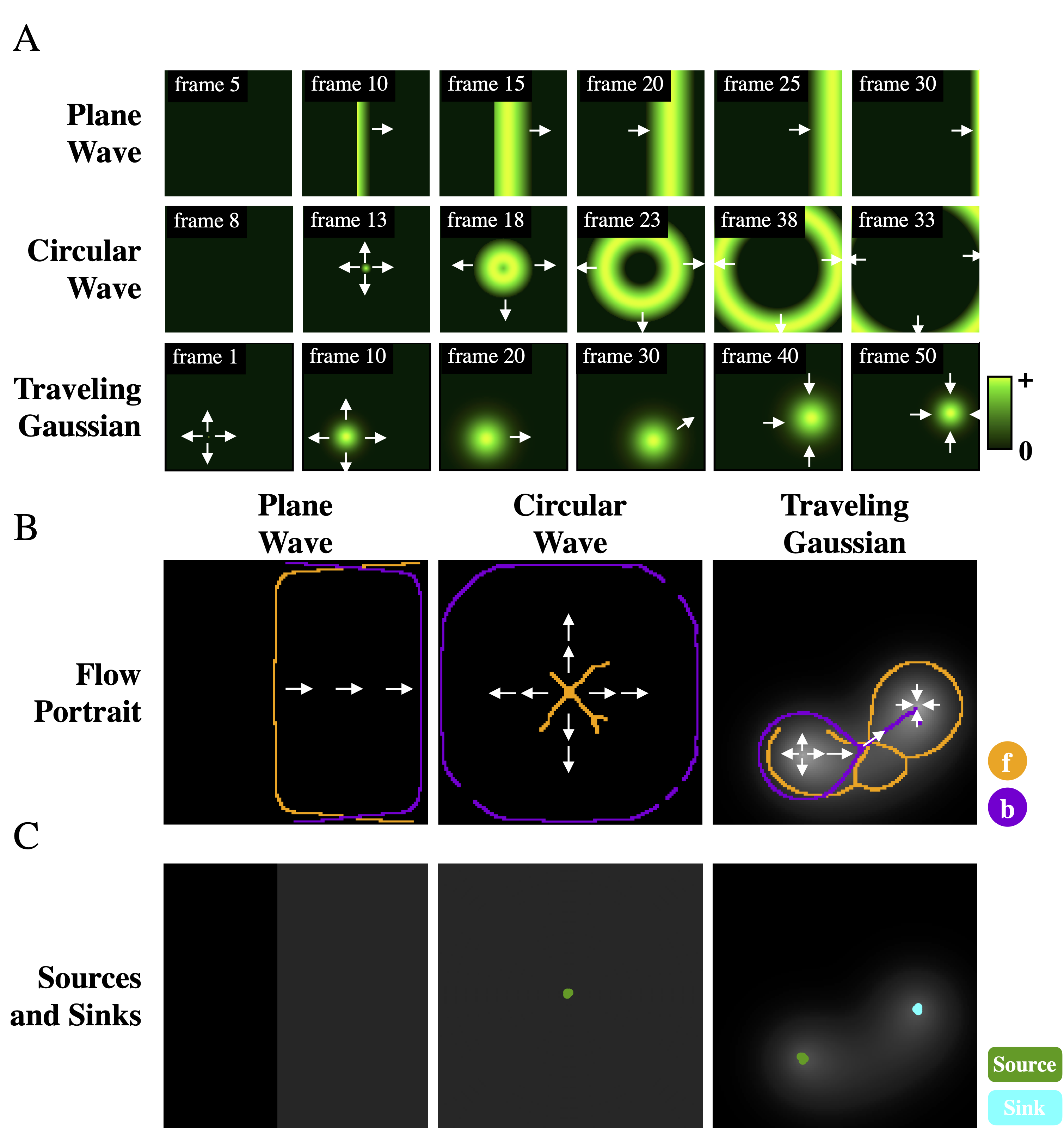}
    \caption{(\textbf{A}) Three additional synthetic examples for which we illustrate FLOW portraits and compare to source sink analysis.
    The 'Plane Wave' example shows a traveling plane wave which begins in the center of the frame and propagates to the right.
    The 'Circular Wave' example shows a traveling circular wave which begins in the center of the frame and expands outwards.
    The 'Traveling Gaussian' example shows a 2-D Gaussian which grows (a source), translates, a shrinks (a sink).
    White arrows indicate the direction of activity propagation.
   (\textbf{B}) The FLOW portraits for both synthetic examples.
   White arrows show the general directions of activity propagation.
   FLOW portraits were computed with integration lengths of 15 frames, 12 frames and 10 frames and the threshold percentile was set to the 91st, 93rd and the 85th percentiles for the plane wave, circular wave and traveling Gaussian datasets respectively.
   (\textbf{C}) Sources (green) and sinks (cyan) identified using the Optical Flow Analysis Toolbox for Wide-field Neuroimaging~\cite{afrashteh2017optical}.}
    \label{synthetic}
\end{figure}

\begin{figure}[htp]
    \centering
    \includegraphics[width=\textwidth]{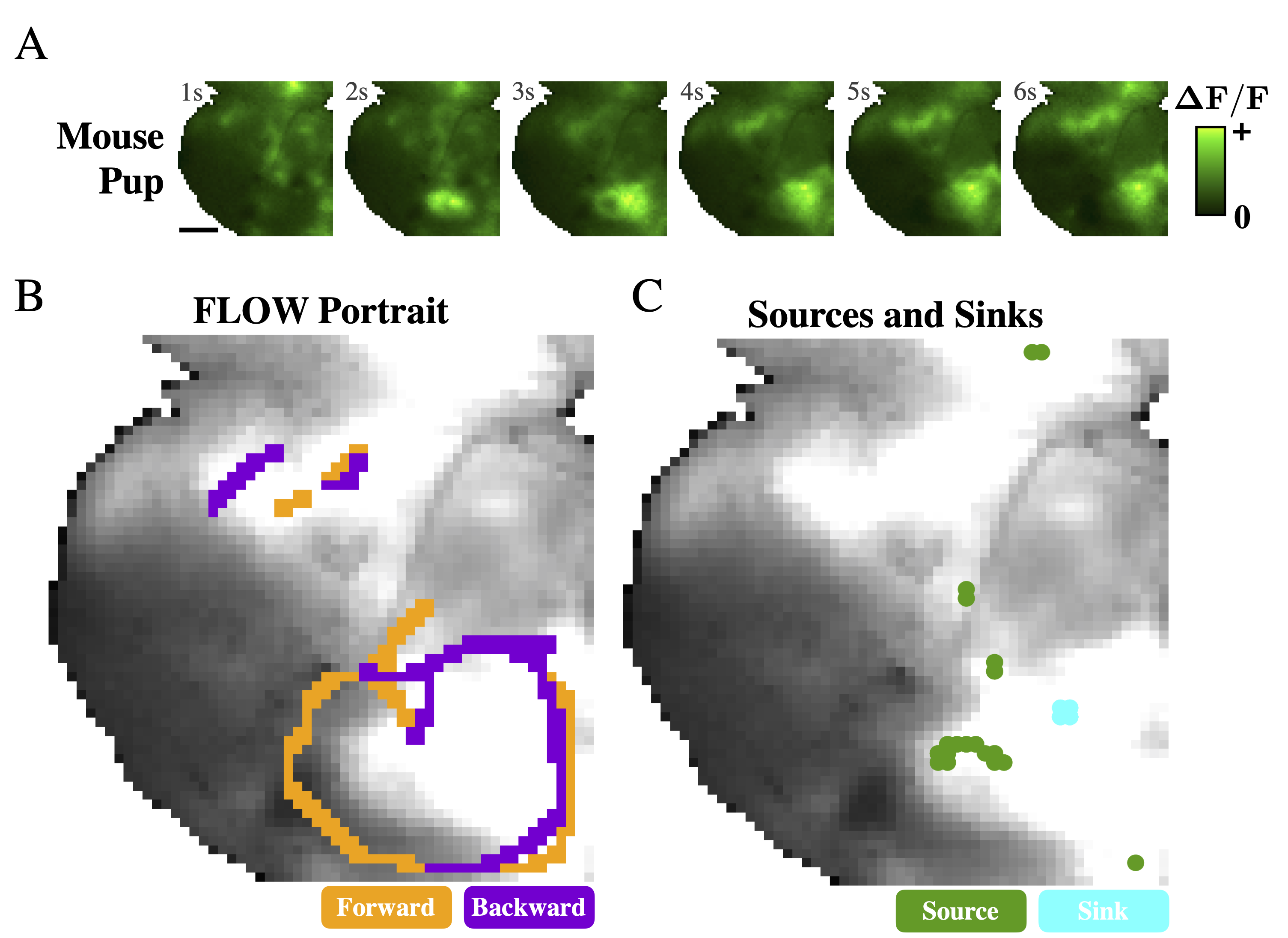}
    \caption{FLOW portraits provide a summary of the flow beyond identifying sources and sinks.
    (\textbf{A}) Widefield imaging data from which FLOW portraits were compared to source sink analysis.
    The widefield imaging dataset depicts a prominent wave towards the bottom of the frame and an additional wave near the top (scale bar is 1mm).
    (\textbf{B}) FLOW portraits for the data shown above with an 20 frame integration length and a $91\%$ threshold percentile.
    (\textbf{C}) Sources (green) and sinks (cyan) identified using the Optical Flow Analysis Toolbox for Wide-field Neuroimaging~\cite{afrashteh2017optical}.
    Comparison with the FLOW portraits highlights how FLOW portraits summarize the overall pattern of activity beyond identifying the sources and sinks.
    }
    \label{ofamm_vs_flow_pup}
\end{figure}

\begin{figure}[htp]
    \centering
    \includegraphics[width=\textwidth]{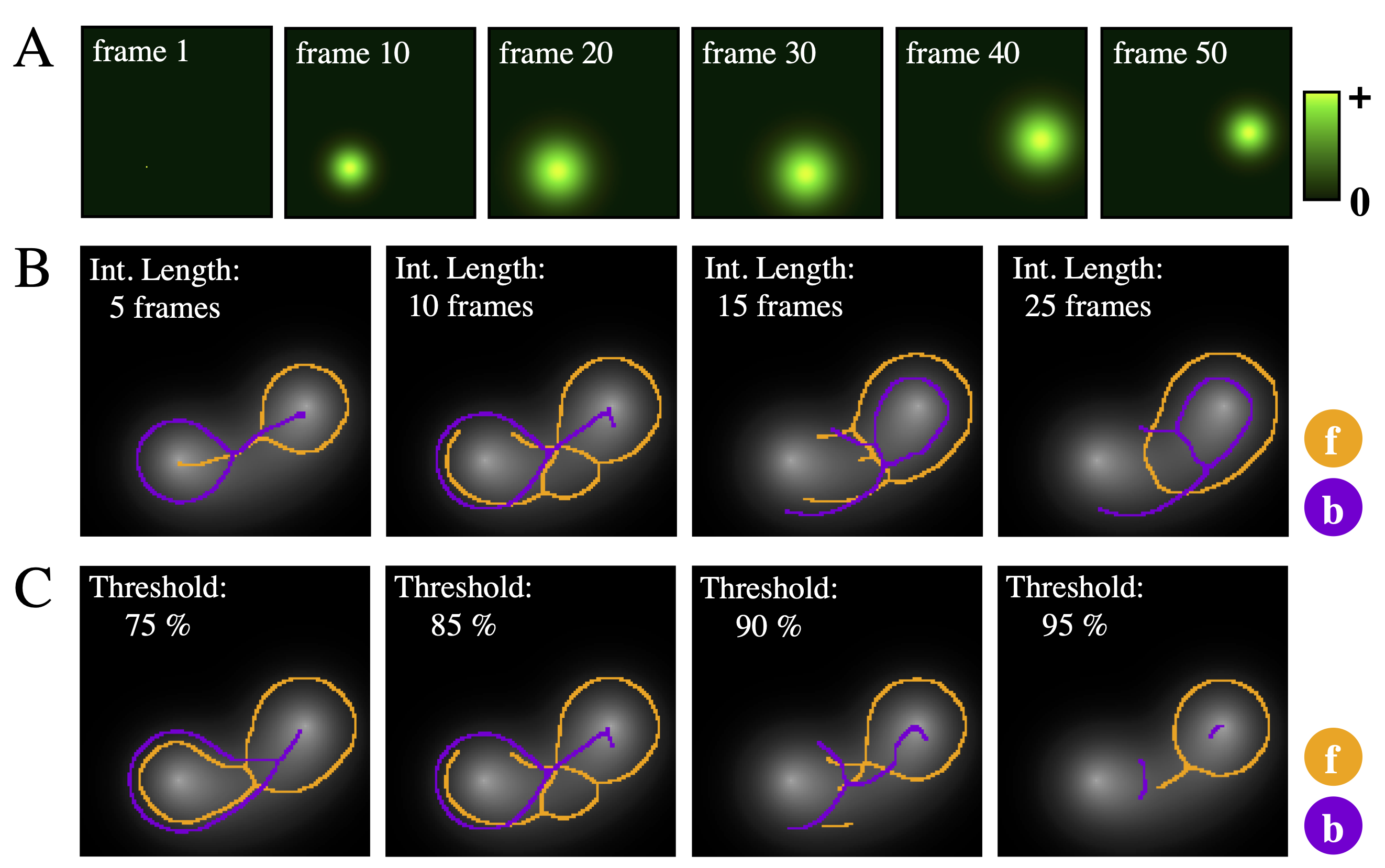}
    \caption{Effects of integration length and threshold percentile hyperparameters on FLOW portraits.
    (\textbf{A}) Synthetic traveling Gaussian examples for which we illustrate the effects of hyperparameters on FLOW portraits.
    (\textbf{B}) Increasing the integration length resolves more detail in the FLOW portrait until a critical integration length is reached. Integration lengths beyond the critical value begin to lose detail. The critical integration length is around 10 frames in this example. All FLOW portraits were computed with the threshold percentile fixed to $85 \%$.
    (\textbf{C}) Thresholding the FTLE field enables isolation of FTLE ridges. However, ridge information will be lost if the threshold percentile is too large. All FLOW portraits were computed with the integration length fixed to $10$ frames.}
    \label{hyperparams}
\end{figure}

\begin{figure}[h!]
    \centering
    \includegraphics[width=\textwidth]{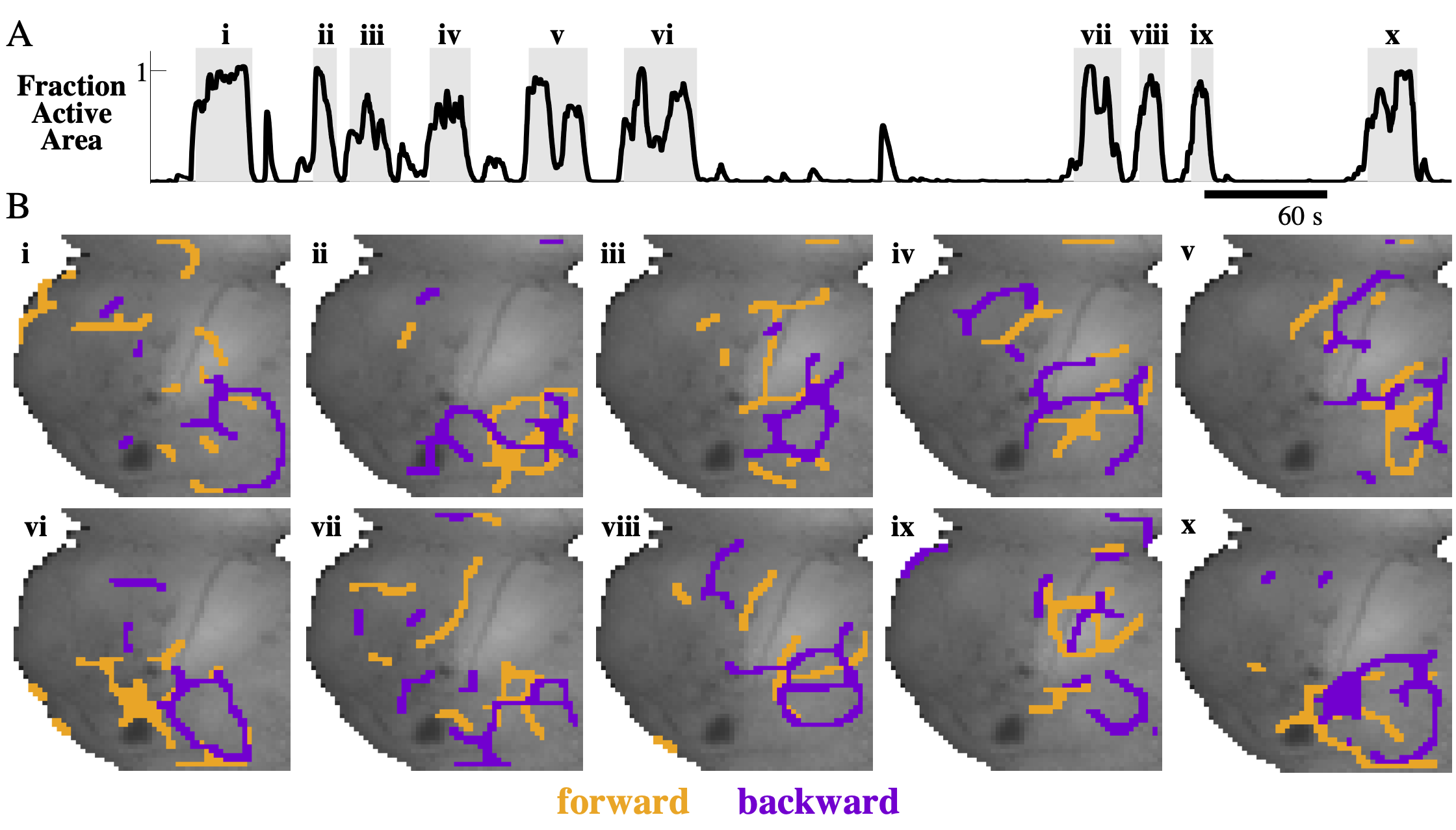}
    \caption{Flow portraits for all examples of pan-cortical waves events analyzed in Figure 5. (\textbf{A}) Ten additional pan-cortical events are seen, as indicated by events where the area of active cortex crosses the $50 \% $ threshold (\textbf{i-x}).
    Events \textbf{iv} and \textbf{x} are analyzed in Figure 5 and are repeated here.
    (\textbf{B}) FLOW portraits for each pan-cortical wave can be used to summarize the distinct activity that occurs during each event.
    All portraits were computed with an integration length of 40 frames and the threshold percentile was set to the 92nd percentile.}
    \label{moody_waves}
\end{figure}

\begin{figure}[hbp!]
    \centering
    \includegraphics[width=\textwidth]{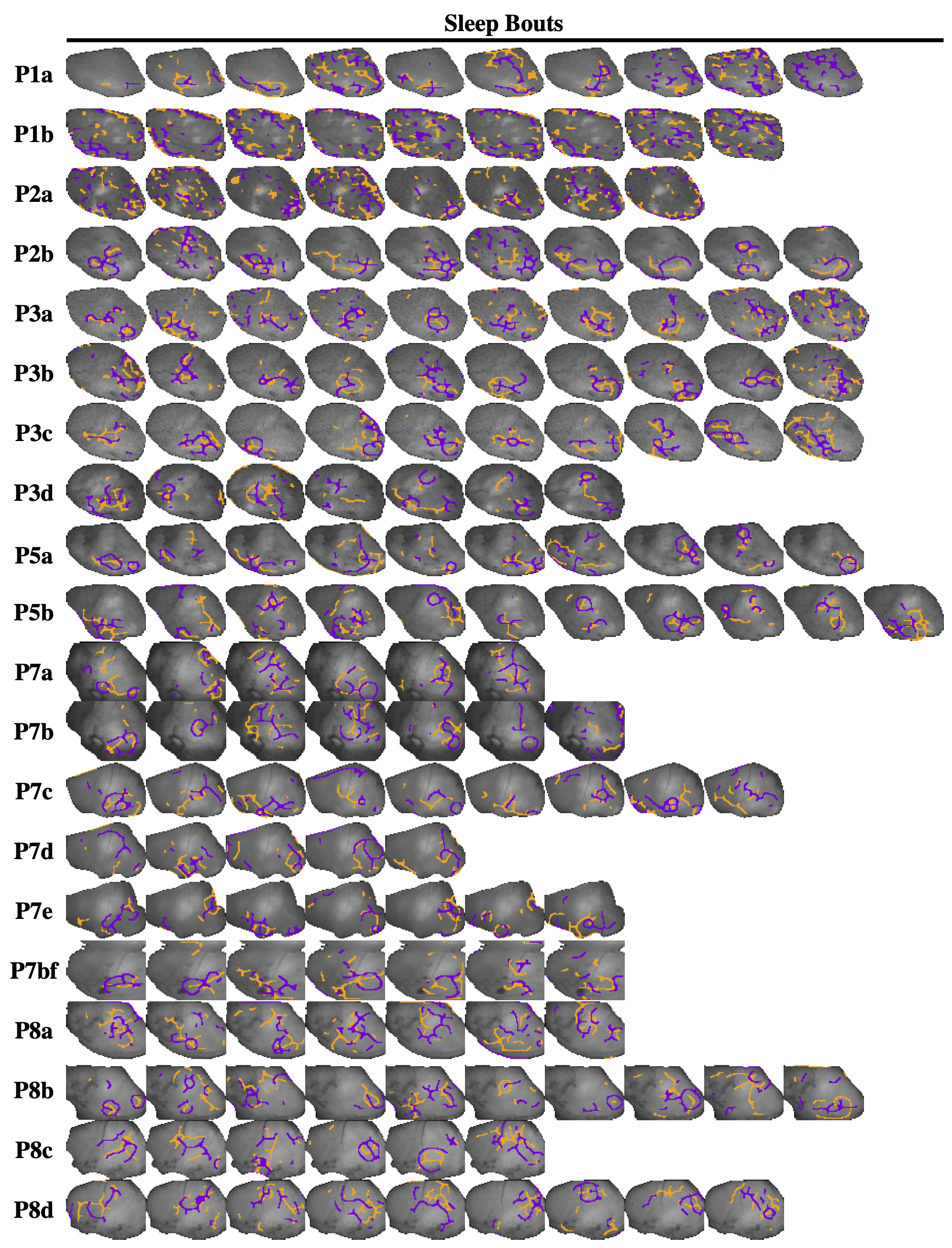}
    \caption{FLOW portraits of all sleep bouts from 20 P1-8 mouse pups are shown.
    A subset of examples from \textbf{P1-8a,b} are shown in Figure 6, while additional examples are shown here.
    All portraits were computed with an integration length of 40 frames and the threshold percentile was set to the 93rd percentile.}
    \label{moody_sleep}
\end{figure}

\clearpage
 \begin{spacing}{.9}
 \small{
 \setlength{\bibsep}{6.5pt}
 \bibliographystyle{IEEEtran}
 \bibliography{supplement}
 }
 \end{spacing}